\documentclass[11pt,preprint]{aastex}

\def\simlt{\lower.5ex\hbox{$\; \buildrel < \over \sim \;$}}

\usepackage{graphicx}
\usepackage{natbib}
\usepackage{epstopdf}
\usepackage{epsfig}
\usepackage{subfig}
\usepackage{color}
\usepackage{datetime}
\usepackage{bbding}

\begin{document}

\title{Spectropolarimetry of Radio-Selected Broad Absorption Line Quasars}
\author{M. A. DiPompeo\altaffilmark{1}, M. S. Brotherton\altaffilmark{1}, R. H. Becker\altaffilmark{2,3}, H.D. Tran\altaffilmark{4}, M.D. Gregg\altaffilmark{2,3}, R. L. White\altaffilmark{5}, S. A. Laurent-Muehleisen\altaffilmark{6}}

\altaffiltext{1}{University of Wyoming, Dept. of Physics and Astronomy 3905, 1000 E. University, Laramie, WY 82071}
\altaffiltext{2}{Institute of Geophysics and Planetary Physics, Lawrence Livermore National Laboratory, 7000 East Avenue, P.O. Box 808, L413, Livermore, CA 94550}
\altaffiltext{3}{University of California, Department of Physics, One Shields Ave., Davis, CA 95616}
\altaffiltext{4}{W.M. Keck Observatory, 65-1120 Mamalahoa Hwy, Kamuela, HI 96743}
\altaffiltext{5}{Space Telescope Science Institute, 3700 San Martin Drive, Baltimore, MD 21218}
\altaffiltext{6}{Illinois Institute of Technology, 3101 South Dearborn St., Chicago, IL 60616}

\begin{abstract}
We report spectropolarimetry of 30 radio-selected broad absorption line (BAL) quasars with the Keck Observatory, 25 from the sample of Becker et al. (2000).  Both high and low-ionization BAL quasars are represented, with redshifts ranging from 0.5 to 2.5.  The spectropolarimetric properties of radio-selected BAL quasars are very similar to those of radio-quiet BAL quasars:  a sizeable fraction (20\%) show large continuum polarization (2-10\%) usually rising toward short wavelengths, emission lines are typically less polarized than the continuum, and absorption line troughs often show large polarization jumps.   There are no significant correlations between polarization properties and radio properties, including those indicative of system orientation, suggesting that BAL quasars are not simply normal quasars seen from an edge-on perspective.
\end{abstract}


\section{INTRODUCTION}

Approximately 10\% of optically selected quasar samples show blueshifted, broad absorption lines (BALs) in their spectra (e.g. Weymann et al. 1991).  These features indicate the presence of massive high-velocity (several percent of light speed) outflows that may have important effects (Vernaleo \& Reynolds 2006, Hopkins et al. 2006, Scannapieco \& Oh 2004); for example, they may enrich the ISM of the quasar host galaxy and surrounding IGM and carry away angular momentum from the system.  They may also contribute to a variety of feedback effects such as regulating host galaxy star formation rates (Hopkins \& Elvis 2010), and limit quasar lifetimes by removing fuel from the nuclear regions (Silk \& Rees 1998, King 2003).  However, the true nature of these objects is not yet fully understood and the question still remains: ``Why do some quasars show BALs while others do not?"

BAL quasars (BALQSOs) of a variety of types have been identified, leading to several sub-classifications.  The majority fall into the category of high ionization BALQSOs (HiBALs), in which the BALs are formed by highly ionized species such as \ion{C}{4} ($\lambda$1549\AA), \ion{Si}{4} ($\lambda$1397\AA) and \ion{N}{5} ($\lambda$1240\AA).  A smaller fraction, on the order of 1\% (optically selected), are LoBALs that also have broad absorption from elements with lower ionizations, such as \ion{Mg}{2} ($\lambda$2799\AA)(Trump et al. 2006).  Even more rare are LoBALs which also show absorption from \ion{Fe}{2} ($\lambda$2380\AA, $\lambda$2600\AA, $\lambda$2750\AA), known as the FeLoBALs (Trump et al. 2006).  All of these subclasses are represented in our sample.

The true fraction of BALQSOs however is not yet well known due to several selection effects and their relative rarity.  Especially for objects with large BAL equivalent widths, the amount of absorption may be significant enough to under represent BALQSOs in magnitude-limited samples.  This is heavily dependent on redshift and the wavelength coverage of the observations as much of the absorption occurs at rest-frame UV and optical wavelengths.  It is possible that this is even more important for extreme LoBALs because of the large number of absorption troughs present shortward of 2800\AA\ (Hewett \& Foltz 2003).  Hewett \& Foltz (2003) use an empirical k-correction to account for differences in spectral energy distributions between BALQSOs and non-BALQSOs, resulting in a higher intrinsic BALQSO fraction of around 22\%.  In a recent catalog of BALQSOs from the Sloan Digital Sky Survey Data Release 5 (SDSS DR5), Gibson et al. (2009) find a BAL fraction of about 15\%, which increases to about 19\% when the fact that BALQSOs tend to be more reddened than non-BALQSOs is accounted for.  Additionally, in a sample of gravitationally lensed quasars, Chartas (2000) finds 7 of 20 (35\%) show BALs.  It has become increasingly apparent that the true fraction of BALQSOs is higher than simple observed fractions.  

BALQSOs are generally a highly polarized subclass of quasars, with 9 of 28 showing optical broadband polarization of 2\% or greater compared to 2 of 115 non-BALQSOs (Hines \& Schmidt 1997).  There have been a few spectropolarimetry surveys done in the last several years, for example Ogle et al. (1999) and Schmidt, Hines, and Smith (1997).  Spectropolarimetry of small samples and individual objects has been published as well covering a wide range of BALQSO types (Goodrich \& Miller 1995, Cohen et al. 1995, Glenn et al. 1994).  In general, these results show common trends among polarized BALQSO properties: continuum polarization can exceed 10\% and the polarization generally decreases at longer wavelengths.  Trends are also seen in the polarization levels of spectral lines; levels in absorption lines tend to be equal to or greater than that of the continuum, and emission lines are generally similarly or less polarized than the continuum.

One proposed explanation for the difference between BALQSOs and normal quasars is a simple orientation scheme.  In this paradigm outflows are common to all quasars, but only approximately 10\% are seen along a line of sight that intersects the outflow and therefore show BALs.  Drawing on the unification of Seyfert galaxies (Antonucci 1993), BALQSOs in this paradigm are those quasars seen at high inclinations with the line of sight cutting through an equatorial outflow.  Weymann et al. (1991) observed that the emission-line properties of BALQSOs are very similar to those of non-BALQSOs and this is taken to support this view, despite the fact that some differences in the spectra are noted as well (see also Turnshek et al. 1997).  The distinctions between BALQSO subclasses have also been explained via orientation; in this scheme LoBALs are those BALQSOs seen at the largest inclinations, where the line of sight begins to cut through the outflow at a larger distance from the central engine so that lower ionization levels are seen.

Spectropolarimetry results have also been used to support the orientation scheme (Goodrich \& Miller 1995, Hines \& Wills 1995, Cohen et al. 1995 for starters).  The interpretation of spectropolarimetry results also drew heavily on analogies with the unification of Seyfert galaxies by orientation.  However, there has been growing evidence that the orientation paradigm is incorrect, at least in its simplest forms.  

Once thought to be strictly radio-quiet (Stocke et al. 1984), relatively recent large and deep radio surveys like FIRST (Becker et al. 1995) and NVSS (Condon et al. 1998) helped show that radio-loud BALQSOs do exist (Brotherton et al. 1998, for example).   Becker et al. (2000) pointed out that BALQSOs exhibit both flat and steep radio spectra, which suggests a much wider range of orientations than just edge on.  Pole-on sources generally show optically thick, beamed synchrotron emission with a flat radio spectrum, while edge-on sources most strongly show optically thin radio lobe emission with a steep radio spectrum.  Zhou et al. (2006) and Ghosh \& Punsly (2007) have also found evidence for short timescale variation of BALQSO radio sources, which is best explained by relativistic beaming and a pole-on orientation, indicating the presence of polar rather than equatorial outflows.

An alternative to the orientation paradigm is a unification by time in which BALQSOs are a short phase in the lifetime of all quasars, and radio observations strengthen this case.  It has been noted that only a small fraction, about 10\%, of BALQSOs are extended at FIRST resolutions (5 arcsec) (Becker et. al. 2000) while about 50\% of non-BALQSOs selected in a similar way are extended sources.  Montenegro-Montes et al. (2008) find in their study of a sample of 15 BALQSOs that all 15 show compact morphology at FIRST resolutions.  Even observations at higher frequencies, thus higher resolutions, show that the majority are very compact, with linear size scales of less than 1 kpc.  High resolution VLBI observations of BALQSOs find similar properties (Jiang \& Wang 2003, Doi et al. 2009).  This size scale along with the convex shape of the radio spectra is typical of compact steep spectrum (CSS) sources, which O'Dea (1998) has suggested are in fact young radio sources.  In addition, Montenegro-Montes et al. (2008) find that the radio spectral index distribution for their sample is not significantly different from a compact non-BALQSO sample, indicating that the BALQSOs do not have a preferred orientation.  Gregg et al. (2002, 2006) suggest that a BAL phase evolves into a radio-loud normal quasar phase with a short overlap.  X-ray observations show that radio-loud BALQSOs tend to be X-ray brighter than radio-quiet BALQSOs, but faint compared to non-BALQSOs (Brotherton et al. 2005, Miller et al. 2009).  This provides more evidence that radio-loud BALQSOs are an intermediate lifetime phase.

Polarization of radio-loud BALQSOs has been somewhat addressed in the past (Wills et al. 1999, Hutsemekers \& Lamy 2000), but a deep look at a large sample is still needed. Using a large sample of BALQSOs with radio information and spectropolarimetry, we can test the orientation schemes.  In order for a measurable polarization signal to be observed, the scattering region cannot be symmetric and therefore there is geometric information present.  Using radio spectral indices and a large sample to ensure statistical significance, we can search for correlations between viewing angle and polarization properties, mainly polarization percentages and polarization position angles.  Combining this with polarization information from various spectral features (continuum, BALs, emission lines) we can potentially reveal the geometry of the central regions of BALQSOs.  To this end we have begun work on a project involving a large sample of BALQSOs for which we have or will be obtaining radio information and spectropolarimetry.  This paper is the first of several from this project, and our aim here is to present the spectropolarimetry data from our Keck sample, with data taken in July 1999, January 2000 and August 2000, and report the most basic results.

We adopt the cosmology of Spergel et al. (2007) for all calculated properties, with $H_{0}=71$ km/s/Mpc, $\Omega_{M}=0.27$ and $\Omega_{\Lambda}=0.73$.

\section{TARGETS}

Our primary source of radio-selected BALQSOs is the northern installment of
the FIRST Bright Quasar Survey (White et al. 2000; Becker et al. 2000).
The FBQS selects stellar objects also detected by the FIRST survey 
(Becker et al. 1995), a deep radio survey at 20 cm, down to a limiting 
$E$\ magnitude (17.8) and a weak color criterion of $O-E<2$ (E and O magnitudes are from the red and blue filters of the Palomar Observatory Sky Survey).  Out of some 
600 quasars reported by White et al. (2000), 29 were identified as BALQSOs or probable BALQSOs.  The uniform selection of this  sample
permits good statistical comparisons to identify differences in properties
of BALQSOs and the parent population.  We obtained spectropolarimetry
for 25 of the 29 Becker et al. (2000) quasars.

We also present spectropolarimetry of: two BAL quasars from the
southern installment of the FIRST Bright Quasar Survey (Becker et al. 2001),
a BAL quasar originally found in the FBQS that later fell out because
of an adjustment in the selection criterion, the FIRST-selected FR II BAL 
quasar reported by Gregg et al. (2000) and a bright NVSS-selected 
BAL quasar (Brotherton et al. 2001).  We include these here for completeness and ease of reference, for a total
of 30 quasars.

Table~\ref{mainpropstbl} lists the targets with some basic parameters and their radio properties.  Parameters involving the B magnitude use SDSS DR7 g-band magnitudes as a direct estimate of the B magnitude; however not all of the objects are found in SDSS, so for those objects B magnitudes quoted in their respective references are used.  Radio spectral indices are taken from their respective sources and use the convention $f_{\nu}\propto\nu^{\alpha}$.  Radio luminosities are 5GHz rest-frame luminosities, calculated using the spectral indices given in the table.  The radio-loudness parameter, log$R^{\ast}$, is the ratio of the 5GHz flux density to the 2500\AA\ optical flux in the rest frame of the object (Stocke et al. 1992), using the radio spectral indices from the table and assuming the optical spectral index $\alpha_{opt}=-1$.  When no radio spectral index is available, it is assumed that $\alpha_r=-0.3$.

\section{SPECTROPOLARIMETRY}
\subsection{Observations}

All observations were made with LRIS, the Low Resolution
Imaging Spectrometer (Oke et al. 1995), in spectropolarimetry mode
(Good\-rich, Cohen, \& Putney 1995; Cohen et al. 1997) on the 10 meter Keck
I (April 2000) and II (July 1999; January 2000) telescopes.  
We used a 300 line mm$^{-1}$ grating blazed at 5000 \AA,
that, with the 1 $\arcsec$ slit (at or near the parallactic angle),
gave an effective resolution of 10 \AA\ (FWHM of lamp lines);
the dispersion was 2.5 \AA\ pixel$^{-1}$.  Wavelength coverage was
typically 3800-8800 \AA.  Observations are broken into four exposures, one for 
each waveplate position (0$\arcdeg$, 45$\arcdeg$, 22.5$\arcdeg$, 67.5$\arcdeg$).
The red end of the spectrum ($\lambda_{obs} >$ 7400\AA) is weakly
contaminated by second-order light at a level of $\simlt$ 5\%
(see e.g., Ogle et al. 1999).  Typical exposure times were on the order of an hour or less, generally 
good enough to obtain detailed information on highly polarized quasars
and make good broad-band measurements of the lower polarization sources.

\begin{figure}
 \centering
  \figurenum{1}
   \includegraphics[width=5in]{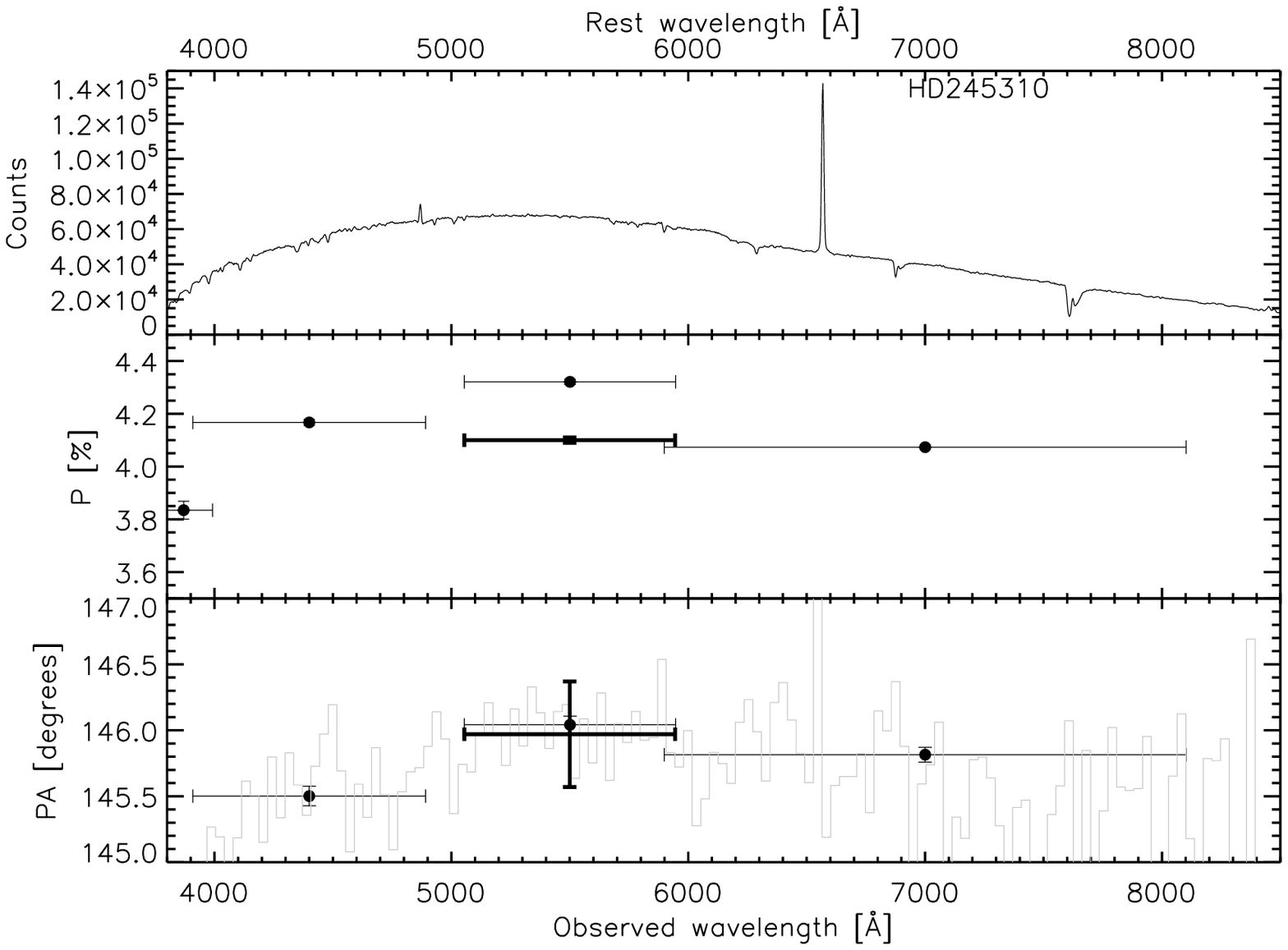}
  \caption{Spectropolarimetry results for the polarization standard star HD245310.  The top panel is the total spectrum (not flux calibrated), the middle panel is the polarization level, and the bottom panel is the polarization position angle.  The polarization and position angle bins are standard UBVR wavelengths.  Bins marked in bold are V-band published values from the literature.  Our values for the polarization P are within about 0.2\%, and the position angle matches to within 0.1$^{\circ}$.\label{stdstarfig}}
\end{figure}

Observations from another Keck program of a bright star through a polaroid filter were used to calibrate the wavelength-dependant polarization position angle changes due to the optics of the instrument.  The polarization efficiency is essentially 1.0 at all wavelengths (Goodrich et al. 1995).  We observed a polarization standard star for each night of observations in order to ensure accurate polarization percentages and to correctly calibrate the zero-point of the polarization position angle.  For the two nights in July 1999 the standard star HD155528 was observed and matched published polarization levels to within a few tenths of a percent, and the position angle was accurate to within $\sim$1$^\circ$.  Standard star HD245310 was observed for the January and April 2000 observations, again matching the published values well.  See Figure~\ref{stdstarfig} for a sample of the standard star results.

\subsection{Measurements}

We reduced our data to one-dimensional spectra using standard techniques within
the IRAF NOAO package. The rms uncertainties in the dispersion solution were
0.2\AA, and we used sky lines to ensure that our zero point was accurate to
0.1\AA.  Wavelengths are air wavelengths.  We followed standard procedures
(Miller, Robinson, \& Goodrich 1988; Cohen et al. 1997) for calculating
Stokes parameters.  Polarizations reported are debiased using the rotated Stokes Q and U parameters, and uncertainties are 1$\sigma$ confidence intervals (Simmons \& Stewart 1985).  Binning is always done in flux before calculating Stokes parameters.

Figure~\ref{polplotsfig} (a-dd) plots our results, showing the total flux spectrum, polarized flux, linear
polarization, and polarization position angle as a function of both
rest-frame and observed-frame wavelengths.  With lower polarization objects requiring heavy binning,
we have attempted to obtain polarization measurements for absorption troughs 
and emission lines using narrower bins.  Continuum polarization measurements were made between the \ion{C}{4} and \ion{C}{3}]/\ion{Al}{3} emission lines where possible, due to the absence of many other features in this range.  For objects with lower redshifts where this was not possible, continuum measurements were made by subjectively choosing the cleanest part of the spectrum.  Table~\ref{contpoltbl} lists the continuum and white-light polarization properties of our sample.  The second to last column indicates the observed-frame wavelength regions where continuum measurements were made; these wavelength bins are plotted in bold in the polarization panel of Figure~\ref{polplotsfig} (a-dd).  The final column, the maximum ISP, is the maximum interstellar polarization possible along the object's line of sight calculated by taking $9\times E(B-V)$ (Serkowsky, Mathewson \& Ford 1975).  The $E(B-V)$ values are from Schlegel et al. (1998) via NED\footnote{This research has made use of the NASA/IPAC Extragalactic Database (NED) which is operated by the Jet Propulsion Laboratory, California Institute of Technology, under contract with the National Aeronautics and Space Administration.}.

For completeness we have also included in table~\ref{othertbl} information for other radio-selected BALQSOs that also have published spectropolarimetry.  All parameters presented there are taken directly from the referenced sources.

To parameterize various properties we make some of the same measurements found in Lamy \& Hutsemekers (2004), as well as a few of our own.  Table~\ref{linepoltbl} lists these measurements.  Average observed polarizations in three main emission features (\ion{C}{4}, \ion{C}{3}] and \ion{Mg}{2}) are given in columns 2, 3, and 4 and are labeled as $p_e$.  Similar measurements were made for the absorption troughs of \ion{C}{4} and \ion{Mg}{2}, given in columns 5 and 6, and are labeled as $p_a$.  The measurements were made by binning across the features, trying to only include the peak of emission lines or the very bottom of absorption troughs.  However, in some instances it was necessary to use wider bins that included the wings of the lines (but not the adjacent continuum) to get a value with reasonable errors.  The last five columns of the table give polarization values normalized by the polarization of the continuum immediately redward of the emission or absorption features in order to account for the fact that the polarization is generally a function of wavelength.  Entries of ``...'' indicate that the feature was not present in the spectrum.  In some cases the signal-to-noise (or the polarization) in the lines was only high enough to obtain upper limits- these values are followed by ($\downarrow$) in the table.  Upper limits were not included in the subsequent analysis.

\subsection{Notes on Individual Objects}

\noindent \textbf{\textit{FBQS1641+3058}}

\noindent This BALQSO is the most polarized object in our sample with $P=7.77\%$ from 6200-6500\AA\ (observed frame) and 7.79\% in white light.  While the flux in the \ion{C}{4} trough bottoms is so low it is hard to analyze the polarization level well there, binning across the whole absorption feature does not indicate any significant rise in polarization.  The polarization in the \ion{Al}{3} absorption line also seems consistent with the continuum.  However there does appear to be a large rise in the \ion{Si}{4} troughs, with a value averaged over all of the features of almost 9\%.  Although close to the red end of the spectrum where contamination from second-order light may be an issue, the polarization in the \ion{Mg}{2} BAL seems to rise over 9\%.  The discrepancy between polarization levels in the \ion{C}{4} absorption troughs and others could be an indication of a higher \ion{C}{4} column density along the polarized line of sight, something not typically seen in these objects.  The emission lines appear polarized at a level very close to that of the continuum.

\noindent \textbf{\textit{FBQS0946+2744}}

\noindent FBQS0946+2744 is likely unpolarized, or only very weakly polarized.  In the wavelength range 5770-6070\AA\ (observed frame) we measure $P=0.18\%$; however, the insterstellar polarization along this line of sight can reach as high as 0.22\%.  Therefore, the polarization is not likely to be intrinsic.  The polarization however may rise to around 0.4\% in the regions around the \ion{N}{5} and Lyman alpha absorption, and there may be some indication of a rotation in the position angle in that region.  This could indicate that there is a small level of intrinsic polarization in this BALQSO, but these observations are not extremely convincing.  The spectrum is also slightly odd, with the absense of any strong emission features even though there is obvious absorption present.

\section{EXTENDED RADIO SOURCES}
Three BALQSOs in our sample have been identified as extended radio sources and their properties are summarized in table~\ref{extendedtbl}.  The criteria for classifying these objects as extended was by comparing the FIRST integrated ($S_i$) and peak ($S_p$) 1.4 GHz fluxes; objects with $S_i-S_p \geq 1$mJy are considered extended.  One of these objects, 1016+5209, is the first FR II BALQSO discovered, from Gregg et al. (2000).  We do note however that upon visual inspection the other two objects appear point-like, but we include them here due to the significant difference in peak and integrated flux.

An interesting comparison one can make for extended sources is between the radio position angle and the polarization position angle, an idea first discussed in Brotherton et al. (1997).  For edge on sources with a polar scattering region, one would see radio and polarization position angles perpendicular to one another- a face on view with an equatorial scattering region will have these two PAs parallel.  Examining this relationship between radio and polarization PAs in BALQSOs then could provide strong evidence either for or against a preferred orientation.  One of the first attempts to make this comparison was in Wills et al. (1999), but the results were inconclusive due to high variability in the polarization properties of the object.  For our sample we are only confident making this comparison with the obviously extended 1016+5209, which has a radio position angle of $146^{\circ}$ and a continuum polarization of $78^{\circ}$.  This is a difference of $68^{\circ}$, so the two are roughly perpendicular.

The fact that only 3 of our 30 objects are resolved (and only one is noticeably visibly extended) at this resolution is consistent with other studies, for example Montenegro-Montes (2008) and Becker et al. (2000) as discussed in \S 1, and may be evidence for the evolution-based models for BALQSOs.

\section{ANALYSIS \& RESULTS}

\begin{figure}
 \centering
  \figurenum{4}
   \includegraphics[angle=90,width=5.5in]{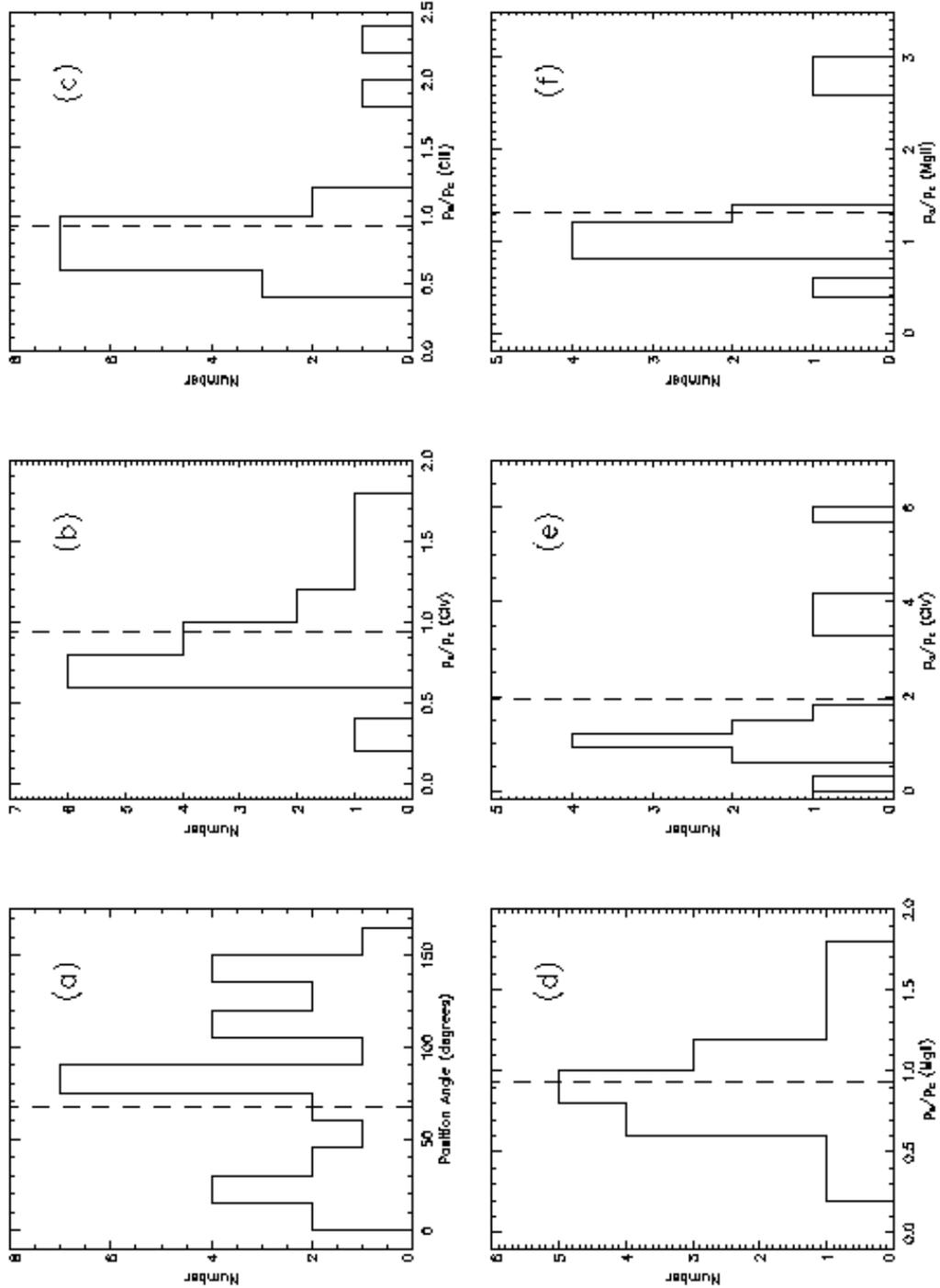}
  \caption{Histograms of the polarization properties (from Table~\ref{contpoltbl} and~\ref{linepoltbl}), except for the continuum polarization level (see Figure~\ref{quietloudfig}).  Box (a) is the polarization position angle, boxes (b), (c) and (d) are the ratios of emission line polarization to the adjacent continuum polarization for \ion{C}{4}, \ion{C}{3}] and \ion{Mg}{2}, respectively.  Boxes (e) and (f) are the ratios of absorption line polarization to the adjacent continuum polarization for \ion{C}{4} and \ion{Mg}{2}, respectively.  Vertical lines indicate the mean of each parameter- the mean of the PA was computed with circular statistics.\label{polhistfig}}
\end{figure}

Figure~\ref{mainhistfig} and figure~\ref{polhistfig} show the distributions of the main properties of the sample, with vertical dashed lines marking the mean of each (the mean of the polarization position angle was calculated using circular statistics).  We can see from these distributions that indeed emission lines are generally polarized at a level similar to or less than the continuum and BALs are generally polarized at a level similar to or greater than the continuum.  The values for $p_e/p_c$ for \ion{C}{4}, \ion{C}{3}] and \ion{Mg}{2} emission lines all average to slightly less than 1 ($0.95$, $0.93$ and $0.94$, respectively).  Values of $p_a/p_c$ for \ion{C}{4} and \ion{Mg}{2} all average to higher than 1 ($1.93$ and $1.30$, respectively).  We do note that given the widths of the distributions for these values, the means are not inconsistent with 1.

We have also done a preliminary search for correlations between numerous properties (most importantly between polarization and radio properties) using the Spearman $r_s$ rank correlation coefficient, and results are shown in table~\ref{corrtbl}.  The number of objects used for each correlation test is given in the final column of the table; given the generally small sample size for each test, we use a loose criterion of $P_{r_s} \le 0.05$ to indicate a significant or marginally significant correlation.   Values of $P_{r_s}$ satisfying this inequality are shown in bold type.  We see that there is no correlation between the continuum polarization and any radio properties, and most importantly there is no correlation with the radio spectral index $\alpha_R$.

\begin{figure}
 \centering
  \figurenum{7}
   \includegraphics[width=5in]{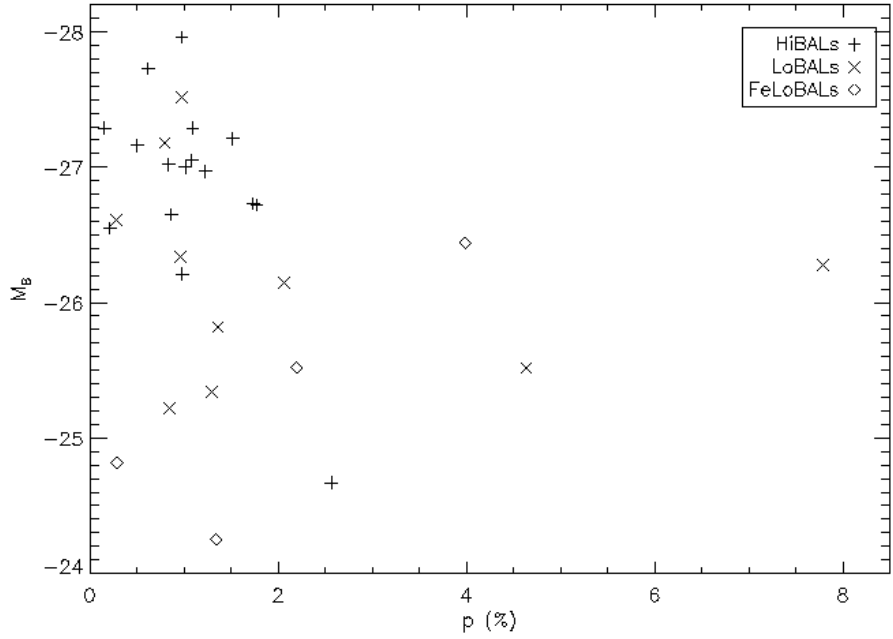}
  \caption{The correlation between continuum polarization and $M_B$.\label{pvsmagfig}}
\end{figure}

\begin{figure}
 \centering
  \figurenum{8}
   \includegraphics[width=5in]{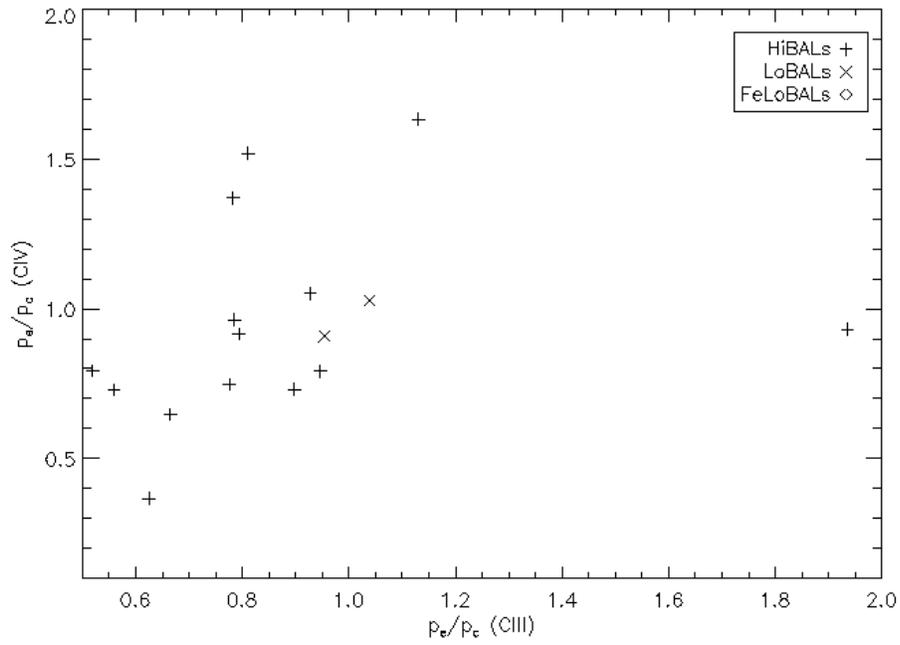}
  \caption{The correlation between polarization in the \ion{C}{4} and \ion{C}{3}] emission lines, each normalized by the respective redward continuum.\label{civciiifig}}
\end{figure}

\begin{figure}
 \centering
  \figurenum{9}
   \includegraphics[width=5in]{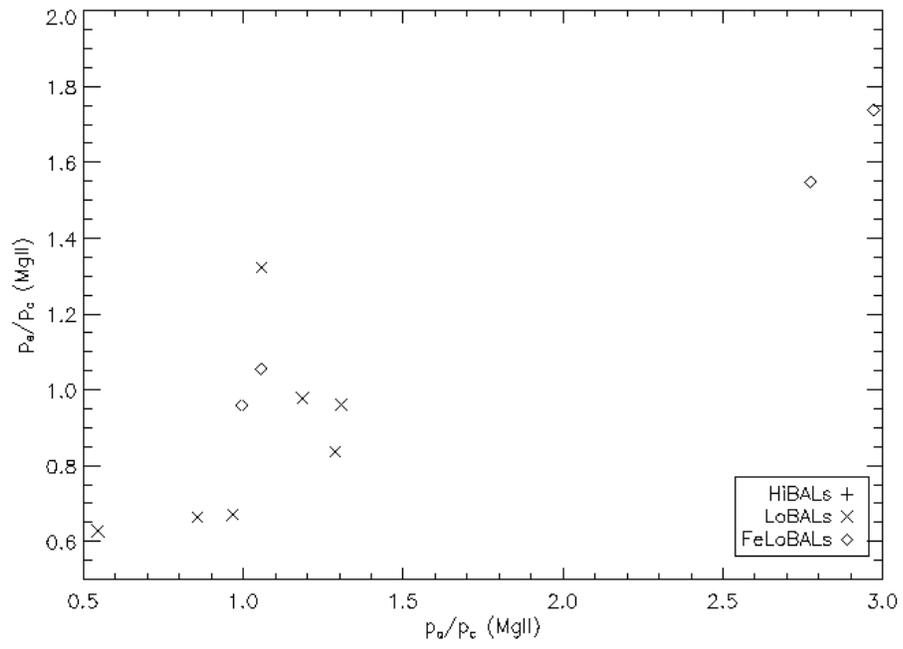}
  \caption{The correlation between polarization in the \ion{Mg}{2} emission line and absorption trough.\label{mgiifig}}
\end{figure}

There were three correlations that satisfied our cut-off (with 19 correlations checked and our chosen cut-off, we would only expect to find about one correlation by chance); continuum polarization and absolute B magnitude, polarization levels in the \ion{C}{4} and \ion{C}{3}] emission lines, and polarization in the \ion{Mg}{2} emission line and absorption line.  All three were positive correlations, and plots of each are shown in figure~\ref{pvsmagfig}, figure~\ref{civciiifig} and figure~\ref{mgiifig}.  Different BAL subclasses are indicated by different symbols in these figures.  The relationship between polarization in the \ion{C}{4} emission line and \ion{Mg}{2} absorption line also meets the criterion, in fact displaying a perfect anti-corrleation; however, only 3 objects had both measurements, so more will be needed to see if this relationship is real.  The correlation between the polarization in the \ion{C}{4} and \ion{C}{3}] emission lines was found by Lamy \& Hutsemekers (2004) and we confirm their results.  However we do not see the same trend of \ion{C}{3}] emission being systematically more polarized than \ion{C}{4} emission.  The relationship between polarization in the \ion{Mg}{2} emission and absorption lines is the strongest correlation, with $P_{r_s}=0.003$.  We will defer discussion of the implications of these correlations to a later paper, which will also include more objects.

\section{DISCUSSION}
\begin{figure}
 \centering
  \figurenum{5}
   \includegraphics[width=5in]{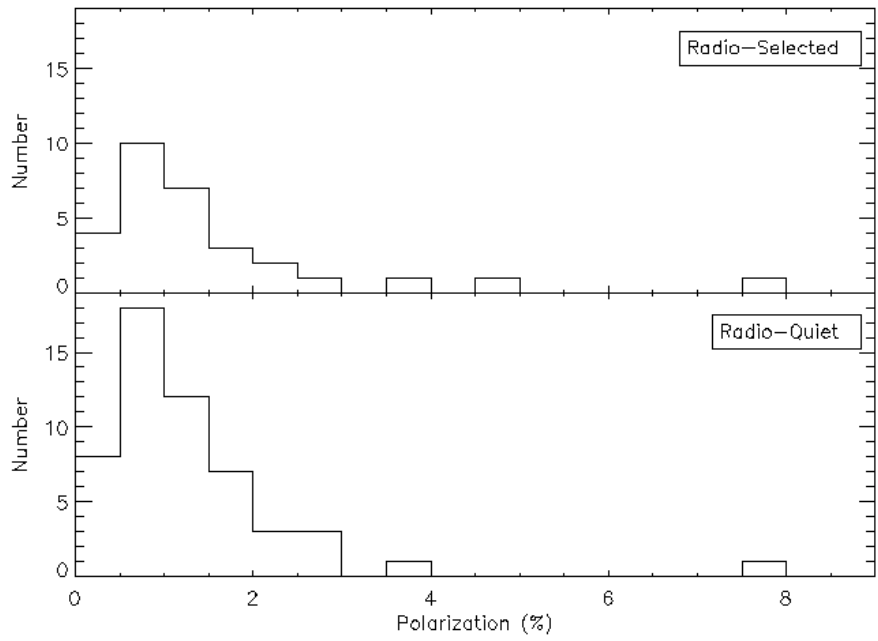}
  \caption{Comparison of the white light polarization levels of our radio-selected BALQSO sample and the radio-quiet sample of Schmidt \& Hines (1999).\label{quietloudfig}}
\end{figure}

The results in the previous section indicate that there is little difference in the polarization properties between radio-selected and radio-quiet BALQSOs.  The rising polarization in BAL troughs and decreasing polarization in emission lines are general trends already seen in radio-quiet BALQSOs, and we see similar trends in our radio-selected sample.  Another similarity is shown in Figure~\ref{quietloudfig}, a histogram of the white-light polarization for our sample versus the radio-quiet sample (53 objects) of Schmidt \& Hines (1999).  The top panel shows our radio-selected sample and the bottom panel is the radio-quiet sample; bin sizes are 0.5\%.  Just through visual inspection it is clear that there is no significant difference between the two; both radio-loud and radio-quiet BALQSOs exhibit similar distributions in polarization.  Though the sample size is relatively small, a simple 2-sample KS test gives $p=0.146$, which offers little reason to believe the distributions are significantly different.  If as a cutoff for defining high broad-band polarization we choose 2\%, we find that 6 of 30 ($20\% \pm 7.3\%$) of our sample are highly polarized and 8 of 53 ($15.1\% \pm  4.9\%$) of the radio-quiet sample are highly polarized.  Both of these are smaller than the 9 of 28 ($32.1\% \pm 8.8\%$) reported by Hines \& Schmidt (1997) (errors are $1\sigma$ values from a binomial distribution).  It should be noted however that the difference between the Schmidt \& Hines (1999) sample and that of Hines \& Schmidt (1997) could be in the choice of cutoff for included objects; the earlier sample only included objects for which $\sigma_p < 0.62\%$, while the later one used a stricter criterion of $\sigma_p < 0.5\%$.

\begin{figure}
 \centering
  \figurenum{6}
   \includegraphics[width=5in]{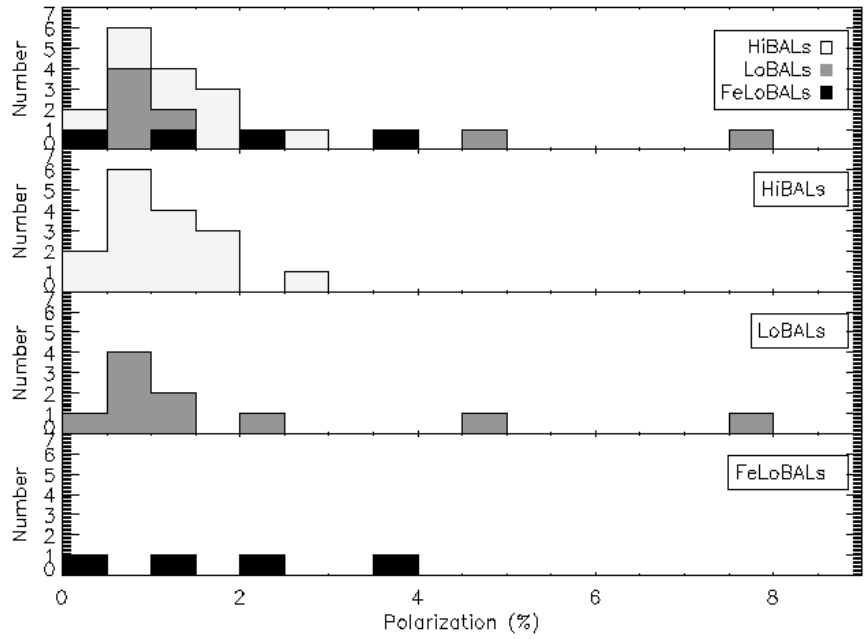}
  \caption{Comparison of the continuum polarization distribution in the three BALQSO subclasses.  The top panel is all three types overlaid on one another (not stacked), while the bottom three panels are the individual distributions shown to make clear any overlaps in the top panel.\label{bytypefig}}
\end{figure}

We have also compared continuum polarization levels between the three BALQSO subtypes (HiBALs, LoBALs and FeLoBALs).  Figure~\ref{bytypefig} shows a comparison of the distributions of each type.  While the numbers are too small to get significant statistics (there are only four FeLoBALs and nine LoBALs), we can see that LoBALs seem to extend to higher polarization levels.  The mean HiBAL polarization about 1.1\%, compared to 2.1\% for LoBALs and 1.9\% for FeLoBALs.  These results are in agreement with the findings of Hutsemekers et al. (1998), in that the highest polarizations are seen in the LoBAL/FeLoBAL subclasses and the range in polarization for those classes is higher than for HiBALs.  Implications of these results will be analyzed when we have added more data to the sample.

The preliminary correlation analysis does not lend any support to the orientation dependent schemes for explaining BALQSOs.  The most important comparison here is between polarization and the radio spectral index.  Since $\alpha_R$ is a rough indicator of orientation (see \S1), the fact that there is no correlation with the polarization may be an indication that BALQSOs do not differ from other QSOs simply by viewing angle.  In the edge-on picture, one would expect the polarization to rise as viewing angle increases.  The radio spectrum should also become steeper as viewing angle increases, and we would thus expect to see a correlation between higher polarization and steeper radio spectrum.  It is possible that because spectral index is only a statistical measure of orientation, and not a one-to-one correspondence, that we may see this correlation only with a larger sample size.  We will be examining this in the near future.  We do note that for the one object we have a measured radio PA (\S4) it is roughly perpendicular to the continuum polarization PA, which is consistent with orientation schemes.

\section{SUMMARY}
\begin{figure}
 \centering
  \figurenum{3}
   \includegraphics[angle=90,width=5.5in]{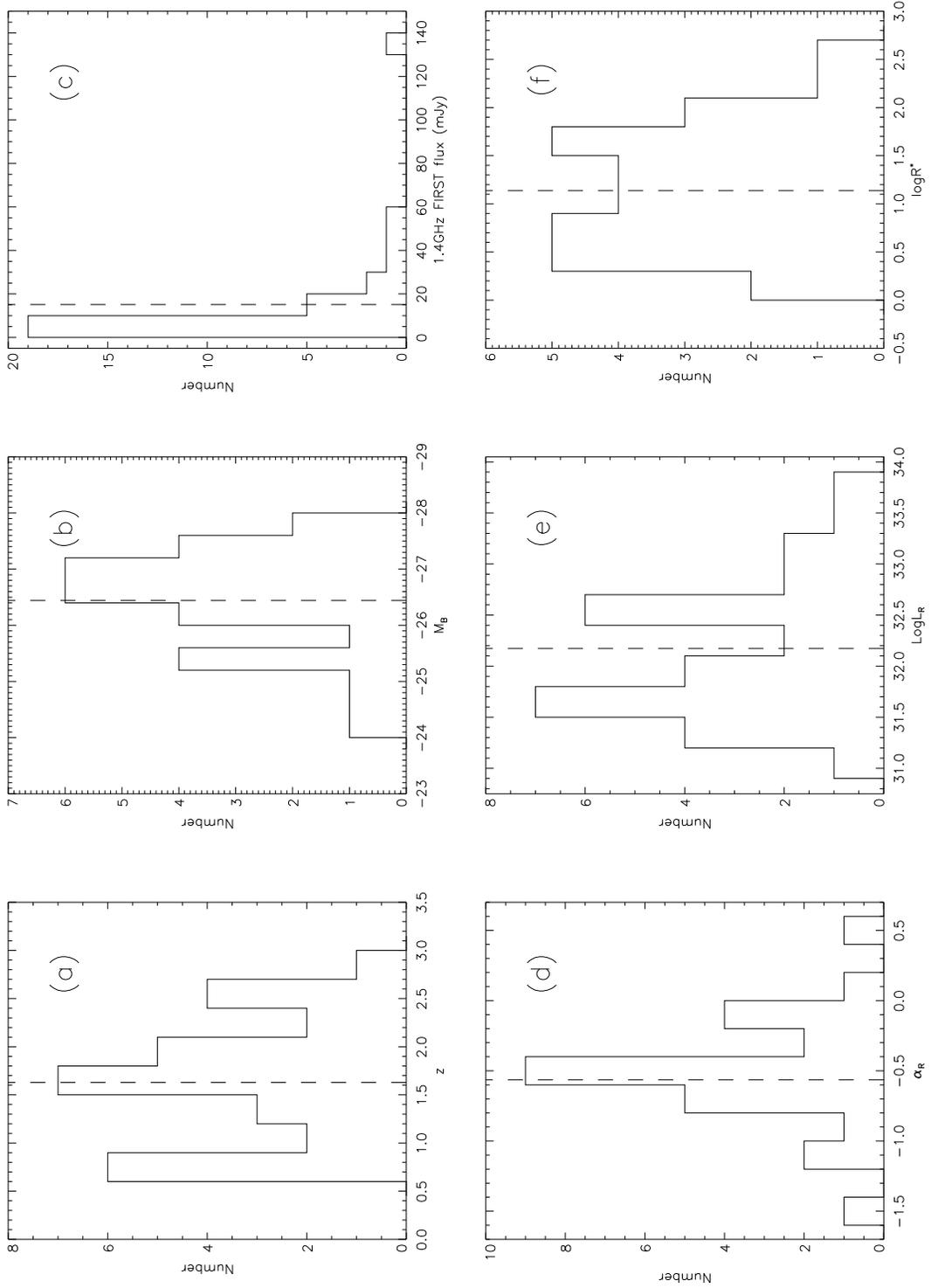}
  \caption{Histograms of the main sample properties (from Table~\ref{mainpropstbl}).  Box (a) is the redshift, box (b) is the absolute B magnitude, box (c) is the FIRST 1.4 GHz flux, box (d) is the radio spectral index, box (e) is the radio luminosity and box (f) is the radio loudness.  Vertical lines indicate the mean of each parameter.\label{mainhistfig}}
\end{figure}

We present spectropolarimetry between 3800 and 8500\AA\ from the Keck Telescope of 30 radio-selected BAL quasars.  They cover a range of redshifts, from about 0.5 to 2.5, and radio properties (see figure~\ref{mainhistfig}).  The various BALQSO subclasses (HiBALs, LoBALs, FeLoBALs) are all represented.  We find the same range of spectropolarimetric properties that are seen in radio-quiet BALQSOs; in general polarization in BAL troughs is higher or similiar to the continuum, emission lines are polarized at a level less than or similar to the continuum, and polarization increases toward shorter wavelengths.  The distributions of white-light polarization for radio-selected and radio-quiet BALQSOs are very similar, and no correlations between radio properties and polarization properties are found.  Three significant/marginally significant correlations are identified and are between continuum polarization and absolute B magnitude, polarization levels in the \ion{C}{4} and \ion{C}{3}] emission lines, and polarization in the \ion{Mg}{2} emission line and absorption line.

Subsequent papers will analyze these findings further in the context of testing the orientation paradigm for BALQSOs.

\acknowledgments

The W. M. Keck Observatory is a scientific partnership 
between the University of California and the California Institute of Technology,
made possible by the generous gift of the W. M. Keck Foundation.
We acknowledge support from NASA through grant \#NNG05GE84G and the Wyoming NASA Space Grant Consortium, NASA Grant \#NNG05G165H.  A portion of this work has been performed under the auspices of the U.S. Department of Energy by Lawrence Livermore National Laboratory under Contract W-7405-ENG-48.

{\it Facilities:} \facility{Keck I (LRISp)}

\clearpage

\begin{deluxetable}{cccccccccc}
 \tabletypesize{\scriptsize}
 \tablewidth{0pt}
 \tablecaption{The radio selected BALQSO sample.\label{mainpropstbl}}
 \tablehead{
  \colhead{RA} & \colhead{DEC} & \colhead{$z$} & \colhead{$S_p$} & \colhead{$S_i$} & \colhead{$M_B$} & \colhead{log$L_r$} & \colhead{log$R^{\ast}$} & \colhead{$\alpha_r$} & \colhead{Class}
  }
   \startdata
   01 35 15.20 & $-$02 13 49.0 &  1.82 & 22.42  & 22.81  & -27.2\tablenotemark{h} & 32.9 & 1.57\tablenotemark{h} & \nodata & HiBAL\tablenotemark{a} \\
   02 56 25.78 & $-$01 19 04.9 &  2.49 & 25.78  & 27.56  & -27.0\tablenotemark{h} & 33.2 & 1.97\tablenotemark{h} & \nodata & HiBAL\tablenotemark{a} \\
   07 24 17.63 & +41 59 18.6   &  1.55 & 7.89   & 7.90   & -25.8 & 32.4 & 1.59 & +0.0    & LoBAL \\
   07 28 30.80 & +40 26 22.0   &  0.66 & 16.96  & 16.79  & -27.5 & 31.7 & 0.25 & -1.1    & LoBAL \\
   08 09 01.25 & +27 53 54.8   &  1.51 & 1.17   & 1.67   & -26.7 & 31.5 & 0.34 & \nodata   & HiBAL \\
   09 10 44.68 & +26 13 01.7   &  1.92 & 7.84   & 7.46   & -26.2 & 32.5 & 1.54 & -0.5    & HiBAL \\
   09 13 29.33 & +39 44 42.9   &  1.58 & 2.06   & 2.09   & -26.7 & 31.7 & 0.61 & -0.6    & HiBAL \\
   09 34 04.96 & +31 53 42.4   &  2.42 & 4.68   & 4.41   & -27.7 & 32.5 & 0.91 & -0.2    & HiBAL \\
   09 46 02.66 & +27 44 03.0   &  1.74 & 3.54   & 3.63   & -27.3 & 31.9 & 0.57 & $<$-1.5 & HiBAL \\
   09 57 07.91 & +23 56 20.0   &  1.99 & 136.10 & 140.49 & -26.7 & 33.8 & 2.60 & -0.6    & HiBAL \\
   10 16 12.48 & +52 09 22.6   &  2.46 & 5.28   & 6.47   & -24.7 & 32.5\tablenotemark{e} & 2.18\tablenotemark{e} & -1.0\tablenotemark{g} & HiBAL\tablenotemark{b}\\
   10 31 10.64 & +39 53 21.7   &  1.08 & 2.45   & 2.03   & -25.4 & 31.6 & 0.97 & -0.2   & LoBAL \\
   10 44 59.83 & +36 56 00.6   &  0.70 & 14.61  & 15.00  & -25.5 & 31.9 & 1.23 & -0.5   & FeLoBAL \\
   10 54 27.20 & +25 36 23.9   &  2.40 & 2.99   & 3.02   & -27.2 & 32.3 & 0.92 & -0.5   & LoBAL \\
   11 22 20.49 & +31 24 57.4   &  1.45 & 12.64  & 12.87  & -26.4 & 32.4 & 1.44 & -0.6   & LoBAL \\
   12 00 51.10 & +35 08 36.5   &  1.70 & 2.03   & 1.46   & -28.0 & 31.8 & 0.12 & -0.8   & HiBAL \\
   12 14 42.64 & +28 03 44.6   &  0.70 & 2.61   & 2.90   & -24.8 & 31.1 & 0.67 & -0.8   & FeLoBAL \\
   13 12 13.94 & +23 20 25.9   &  1.51 & 43.27  & 44.12  & -27.2 & 33.0 & 1.65 & -0.8   & HiBAL \\
   14 08 00.45 & +34 51 25.1   &  1.22 & 2.91   & 2.87   & -26.6 & 31.7 & 0.55 & -0.6   & LoBAL \\
   14 08 07.00 & +30 54 39.7   &  0.84 & 3.34   & 3.21   & -25.2 & 31.4 & 0.81 & -0.7   & LoBAL \\
   14 13 34.40 & +42 12 01.7   &  2.81 & 17.79  & 18.74  & -27.0 & 33.2 & 1.88 & -0.2   & HiBAL \\
   14 20 14.03 & +25 33 54.2   &  2.20 & 1.27   & 1.17   & -27.3 & 31.8 & 0.40 & -1.1   & HiBAL \\
   14 27 03.64 & +27 09 40.2   &  1.17 & 2.58   & 2.98   & -24.3 & 31.5 & 1.38 & -0.7   & FeLoBAL \\
   15 23 15.35 & +37 59 20.7   &  1.34 & 1.67   & 1.83   & -26.2 & 31.5 & 0.57 & -0.6   & LoBAL \\
   15 23 50.43 & +39 14 04.8   &  0.66 & 3.75   & 4.07   & -25.5 & 31.3 & 0.63 & -0.4   & LoBAL \\
   16 03 55.09 & +30 02 02.1   &  2.03 & 53.69  & 54.18  & -27.1 & 33.4 & 2.08 & -0.6   & HiBAL \\
   16 41 52.29 & +30 58 51.7   &  2.00 & 2.14   & 2.66   & -26.3 & 32.0 & 1.06 & +0.5   & LoBAL \\
   16 55 44.18 & +39 45 09.2   &  1.75 & 10.15  & 10.16  & -26.6 & 32.6 & 1.47 & -0.2   & HiBAL \\
   17 09 19.90 & +28 18 35.0   &  2.30 & 1.51   & 2.15   & -27.0 & 31.9 & 0.67 & \nodata     & HiBAL\tablenotemark{c} \\
   23 59 52.56 & $-$12 41 37.2 &  0.87 &  \nodata    & 39.5\tablenotemark{f}& -26.4 & 32.6 & 1.52\tablenotemark{h} & -0.36  & FeLoBAL\tablenotemark{d} \\
   \enddata
 \tablecomments{All objects are from the sample of Becker et al. (2000) unless footnoted in the Class column.  $S_{p}$ is the 1.4GHz FIRST peak flux (mJy/beam) and $S_{i}$ is the 1.4GHz integrated FIRST flux (mJy).  The parameter $\alpha_r$ is the radio spectral index between 1.4GHz (20 cm) and 8.3GHz (3.6 cm) unless noted otherwise, and are quoted from their respective sources.  See text for discussion of other parameters.}
 \tablenotetext{a}{From Becker et al. (2001)}
 \tablenotetext{b}{From Gregg et al. (2000)}
 \tablenotetext{c}{Removed from FBQS sample of Becker et al. (2000)}
 \tablenotetext{d}{From Brotherton et al. (2001)}
 \tablenotetext{e}{Using total radio flux from core and lobes}
 \tablenotetext{f}{1.4GHz NVSS integrated flux}
 \tablenotetext{g}{Calculated spectral index between 1.4GHz NVSS flux and 365MHz WENSS flux}
 \tablenotetext{h}{Not in SDSS, B magnitudes used from references}
\end{deluxetable}

\clearpage

\begin{deluxetable}{ccccccc}
 \tabletypesize{\scriptsize}
 \tablewidth{0pt}
 \tablecaption{Continuum and white light polarization properties of the sample.\label{contpoltbl}}
 \tablehead{
   \colhead{Object} & \colhead{White P(\%)\tablenotemark{a}} & \colhead{White PA($^{\circ}$)\tablenotemark{a}} & \colhead{Cont. P(\%)} & \colhead{Cont. PA($^{\circ}$)} & \colhead{Cont. $\lambda$(\AA)} & \colhead{Max. ISP (\%)}
   }
  \startdata
   0135$-$0213 & 1.51$\pm$0.03 & 114$\pm$1 & 1.50$\pm$0.07 & 111$\pm$1 & 4540-5180 & 0.34 \\   
   0256$-$0119 & 1.22$\pm$0.03 & 28$\pm$1  & 1.26$\pm$0.07 & 56$\pm$2  & 5530-6446 & 0.54 \\
   0724+4159   & 1.35$\pm$0.03 & 2$\pm$1   & 1.15$\pm$0.10 & 4$\pm$2   & 5930-6230 & 0.79 \\
   0728+4026   & 0.97$\pm$0.01 & 133$\pm$1 & 1.12$\pm$0.04 & 133$\pm$1  & 5965-6265 & 0.45 \\
   0809+2753   & 1.73$\pm$0.03 & 64$\pm$1  & 1.81$\pm$0.07 & 61$\pm$1  & 3950-4510 & 0.32 \\
   0910+2613   & 0.98$\pm$0.03 & 801$\pm$1  & 0.99$\pm$0.09 & 82$\pm$3  & 6450-6850 & 0.29 \\
   0913+3944   & 0.86$\pm$0.04 & 81$\pm$1  & 0.72$\pm$0.13 & 88$\pm$5  & 6400-6770 & 0.14 \\
   0934+3153   & 0.62$\pm$0.06 & 16$\pm$2  & 0.53$\pm$0.21 & 29$\pm$9  & 5400-6300 & 0.17 \\
   0946+2744   & 0.14$\pm$0.02 & 85$\pm$4  & 0.18$\pm$0.07 & 64$\pm$12 & 5770-6070 & 0.22 \\
   0957+2356   & 1.77$\pm$0.03 & 12$\pm$1  & 1.93$\pm$0.07 & 12$\pm$1  & 4770-5480 & 0.31 \\
   1016+5209   & 2.56$\pm$0.09 & 78$\pm$1  & 2.87$\pm$0.32 & 79$\pm$3   & 5470-6070 & 0.05 \\
   1031+3953   & 1.30$\pm$0.11 & 45$\pm$2  & 1.69$\pm$0.19 & 38$\pm$3  & 4340-5440 & 0.11 \\
   1044+3656   & 2.19$\pm$0.02 & 83$\pm$1  & 2.18$\pm$0.08 & 81$\pm$1  & 5725-6325 & 0.13 \\
   1054+2536   & 0.79$\pm$0.16 & 26$\pm$5  & 0.83$\pm$0.33 & 7$\pm$10  & 5410-6110 & 0.22 \\
   1122+3124   & 0.96$\pm$0.07 & 115$\pm$2 & 0.71$\pm$0.22 & 120$\pm$8 & 5770-6460 & 0.14 \\
   1200+3508   & 0.97$\pm$0.04 & 58$\pm$1  & 1.42$\pm$0.11 & 58$\pm$2  & 4375-4965 & 0.18 \\
   1214+2803   & 0.28$\pm$0.04 & 116$\pm$4 & 0.40$\pm$0.13 & 111$\pm$10 & 6530-6840 & 0.13 \\
   1312+2320   & 0.50$\pm$0.04 & 147$\pm$2 & 1.02$\pm$0.11 & 110$\pm$3 & 4010-4510 & 0.12 \\
   1408+3451   & 0.27$\pm$0.03 & 43$\pm$3  & 0.32$\pm$0.04 & 45$\pm$4  & 4480-5960 & 0.15 \\
   1408+3054   & 0.85$\pm$0.05 & 100$\pm$1 & 0.87$\pm$0.14 & 100$\pm$4 & 6965-7570 & 0.09 \\
   1413+4212   & 0.83$\pm$0.04 & 19$\pm$1  & 1.18$\pm$0.10 & 19$\pm$2  & 6000-6760 & 0.08 \\
   1420+2533   & 1.09$\pm$0.05 & 107$\pm$1 & 0.89$\pm$0.10 & 106$\pm$3 & 5070-5735 & 0.16 \\
   1427+2709   & 1.34$\pm$0.05 & 79$\pm$1  & 2.01$\pm$0.25 & 67$\pm$6  & 4350-4815 & 0.17 \\
   1523+3759   & 2.06$\pm$0.03 & 76$\pm$1  & 2.23$\pm$0.11 & 76$\pm$1  & 4960-5560 & 0.14 \\
   1523+3914   & 4.63$\pm$0.02 & 64$\pm$1  & 4.73$\pm$0.07 & 65$\pm$1  & 5500-5800 & 0.19 \\
   1603+3002   & 1.08$\pm$0.03 & 136$\pm$1 & 1.15$\pm$0.07 & 136$\pm$2 & 4820-5540 & 0.33 \\
   1641+3058   & 7.79$\pm$0.04 & 145$\pm$1 & 7.77$\pm$0.14 & 149$\pm$1 & 6200-6500 & 0.23 \\
   1655+3945   & 0.21$\pm$0.04 & 130$\pm$6 & 0.36$\pm$0.10 & 129$\pm$8 & 4330-5000 & 0.17 \\
   1709+2818   & 1.01$\pm$0.06 & 162$\pm$2 & 1.34$\pm$0.11 & 163$\pm$2 & 5360-6200 & 0.68 \\
   2359$-$1241 & 3.98$\pm$0.01 & 145$\pm$1 & 4.05$\pm$0.06 & 145$\pm$1 & 5570-5870 & 0.26 \\
  \enddata
 \tablecomments{White light polarization measurements are averaged over the whole spectrum, from 3800-8500 \AA.  Continuum polarization measurements are averaged over the (observed) wavelength region listed in column 6.  Maximum interstellar polarizations along the line of sight using the Serkowsky Law are listed in column 7.}
 \tablenotetext{a}{White light polarization measurements are averaged over the whole spectrum, from 3800-8500 \AA}
\end{deluxetable}

\clearpage

\begin{deluxetable}{ccccccccccc}
 \tabletypesize{\scriptsize}
 \rotate
 \tablewidth{0pt}
 \tablecaption{Polarization properties of emission and absorption features.\label{linepoltbl}}
 \tablehead{
    \colhead{Object} & \colhead{$p_e$ (\ion{C}{4})} & \colhead{$p_e$ (\ion{C}{3}])} & \colhead{$p_e$ (\ion{Mg}{2})} & \colhead{$p_a$ (\ion{C}{4})} & \colhead{$p_a$ (\ion{Mg}{2})} & \colhead{$\frac{p_e}{p_c}$ (\ion{C}{4})} & \colhead{$\frac{p_e}{p_c}$ (\ion{C}{3}])} & \colhead{$\frac{p_e}{p_c}$ (\ion{Mg}{2})} & \colhead{$\frac{p_a}{p_c}$ (\ion{C}{4})} & \colhead{$\frac{p_a}{p_c}$ (\ion{Mg}{2})}
  }
   \startdata
   0135$-$0213 & 1.19 & 1.31 & 1.03 & 2.02 &  \nodata  & 0.73 &  0.90 &  0.64 &  1.24 &   \nodata  \\
   0256$-$0119 & 0.83 & 1.07 &  \nodata  & 1.68 &  \nodata  & 0.75 &  0.78 &   \nodata  &  1.51 &   \nodata  \\
   0724+4159   & 1.29 & 1.25 & 1.22 & 5.27 & 1.66 & 0.91 &  0.95 &  0.96 &  3.71 &  1.31 \\
   0728+4026   &  \nodata  &  \nodata  & 0.61 &  \nodata  & 0.88 &  \nodata  &   \nodata  &  0.67 &   \nodata  &  0.97 \\
   0809+2753   & 1.41 & 1.75 & 1.37 & 1.85 &  \nodata  & 0.79 &  0.95 &  0.86 &  1.04 &   \nodata  \\
   0910+2613   &  \nodata  &  \nodata  &  \nodata  & 1.02 &  \nodata  &  \nodata  &   \nodata  &   \nodata  &  0.01 &   \nodata  \\
   0913+3944   & 0.96 & 0.79 & 0.28($\downarrow$) & 4.04 & 0.65 & 1.37 &  0.78 &  0.37($\downarrow$) &  5.77 &  0.87 \\
   0934+3153   & 0.33 & 0.70 &  \nodata  & 0.75 &  \nodata  & 0.37 &  0.63 &   \nodata  &  0.83 &   \nodata  \\
   0946+2744   &  \nodata  &  \nodata  &  \nodata  & 0.32($\downarrow$) &  \nodata  &  \nodata  &   \nodata  &   \nodata  & 1.78($\downarrow$) &   \nodata  \\
   0957+2356   & 1.86 & 1.91 & 0.60 & 1.38($\downarrow$) &  \nodata  & 1.05 &  0.93 &  0.33\tablenotemark{a} &  0.78($\downarrow$) &   \nodata  \\
   1016+5209   & 2.57 & 2.34 &  \nodata  & 3.15 &  \nodata  & 0.96 &  0.79 &   \nodata  &  1.18 &   \nodata  \\
   1031+3953   &  \nodata  & 2.25 & 0.83($\downarrow$) &  \nodata  & 1.44 &  \nodata  &  0.95 &  0.62($\downarrow$) &   \nodata  &  1.08 \\
   1044+3656   &  \nodata  &  \nodata  & 2.47 &  \nodata  & 2.47 &  \nodata  &   \nodata  &  1.06 &   \nodata  &  1.06 \\
   1054+2536   & 1.35($\downarrow$) & 1.26 &  \nodata  & 3.63 &  \nodata  & 1.52($\downarrow$) &  2.38 &   \nodata  &  4.08 &   \nodata  \\
   1122+3124   & 0.66($\downarrow$) & 0.65 & 1.19 &  \nodata  & 0.95 & 0.36($\downarrow$) &  0.48 &  1.32 &   \nodata  &  1.06 \\
   1200+3508   & 1.10 & 0.99 & 0.63 & 1.78 &  \nodata  & 0.65 &  0.66 &  1.13 &  1.05 &   \nodata  \\
   1214+2803   &  \nodata  &  \nodata  & 0.48 &  \nodata  & 0.86 &  \nodata  &   \nodata  &  1.58 &   \nodata  &  2.77 \\
   1312+2320   & 1.07 & 0.60 & 0.92 &  \nodata  &  \nodata  & 0.93 &  1.94 &  1.17 &   \nodata  &   \nodata  \\
   1408+3451   &  \nodata  & 0.39 & 0.13 &  \nodata  & 0.34($\downarrow$) &  \nodata  &  0.66 &  0.41 &   \nodata  &  1.06($\downarrow$) \\
   1408+3054   &  \nodata  &  \nodata  & 0.53 &  \nodata  & 0.46 &  \nodata  &   \nodata  &  0.63 &   \nodata  &  0.54 \\
   1413+4212   & 0.78 & 0.85 &  \nodata  & 0.70($\downarrow$) &  \nodata  & 0.73 &  0.56 &   \nodata  &  0.65($\downarrow$) &   \nodata  \\
   1420+2533   & 1.52 & 0.77 &  \nodata  & 1.45 &  \nodata  & 1.52 &  0.81 &   \nodata  &  1.45 &   \nodata  \\
   1427+2709   &  \nodata  &  \nodata  & 1.20 &  \nodata  & 2.05 &  \nodata  &   \nodata  &  1.74 &   \nodata  &  2.97 \\
   1523+3759   &  \nodata  & 1.80 & 1.74 &  \nodata  & 2.68 &  \nodata  &  0.88 &  0.84 &   \nodata  &  1.29 \\
   1523+3914   &  \nodata  &  \nodata  & 3.14 &  \nodata  & 4.06 &  \nodata  &   \nodata  &  0.66 &   \nodata  &  0.86 \\
   1603+3002   & 0.99 & 1.00 &  \nodata  & 0.66($\downarrow$) &  \nodata  & 0.92 &  0.79 &   \nodata  &  0.61($\downarrow$) &   \nodata  \\
   1641+3058   & 7.81 & 7.87 & 7.48 & 7.31 & 9.07 & 1.03 &  1.04 &  0.98\tablenotemark{a} &  0.96 &  1.19\tablenotemark{a} \\
   1655+3945   & 0.49 & 0.26 & 0.80($\downarrow$) & 1.08 &  \nodata  & 1.63 &  1.13 &  2.96($\downarrow$) &  3.60 &   \nodata  \\
   1709+2818   & 0.96 & 0.42 &  \nodata  & 0.73 &  \nodata  & 0.79 &  0.52 &   \nodata  &  0.60 &   \nodata  \\
   2359$-$1241 &  \nodata  &  \nodata  & 4.01 &  \nodata  & 4.16 &  \nodata  &   \nodata  &  0.96 &   \nodata  &  0.99 \\
   \enddata
  \tablecomments{Typical uncertainties on $p_e$ and $p_a$ are approximately 0.2-0.4\%.  See text for discussion of all parameters.  Values followed by ($\downarrow$) are upper limits.}
  \tablenotetext{a}{The continuum polarization for these objects was measured blueward of the emission lines because the lines are at the red edge of the spectrum.}
\end{deluxetable}      

\clearpage
\begin{deluxetable}{cccccccc}
 \tabletypesize{\scriptsize}
 \tablewidth{0pt}
 \tablecaption{Extended radio sources.\label{extendedtbl}}
 \tablehead{
   \colhead{} &
   \colhead{Deconvolved} &
   \colhead{Deconvolved} &
   \colhead{Deconvolved} &
   \colhead{Measured} &
   \colhead{Measured} &
   \colhead{Measured} \\
   \colhead{Object} &
   \colhead{Major Axis($\arcsec$)} & 
   \colhead{Minor Axis($\arcsec$)} & 
   \colhead{PA($^{\circ}$)} & 
   \colhead{Major Axis($\arcsec$)} & 
   \colhead{Minor Axis($\arcsec$)} &
   \colhead{PA($^{\circ}$)} &
   }
  \startdata
  0256$-$0119 & 1.73 & 1.30 & 45.9    & 6.58 & 5.61 & 3.2         \\
  0957+2356     & 1.11 & 0.81 & 146.5 & 5.51 & 5.46 & 146.5     \\  
  1016+5209     & 3.93 & 0.00 & 165.2 & 6.68 & 5.35 & 165.2     \\
   \enddata
  \tablecomments{Values are taken from the FIRST database.} 
\end{deluxetable}

\clearpage

\begin{deluxetable}{ccccc}
 \tabletypesize{\scriptsize}
 \tablewidth{0pt}
 \tablecaption{Correlation analysis of polarization properties.\label{corrtbl}}
 \tablehead{
   \colhead{Property 1} & \colhead{Property 2} & \colhead{$r_s$} & \colhead{$P_{r_s}$} & \colhead{n}
   }
  \startdata                
  Cont. $p$                          & $\alpha_R$                      &  0.252 & 0.214              & 26 \\
  Cont. $p$                          & $LogL_R$                        & -0.005 & 0.980              & 30 \\
  Cont. $p$                          & $LogR^*$                         &  0.211 & 0.264              & 30 \\
  Cont. $p$                          & $M_B$     	                  &  0.377 & \textbf{0.040} & 30 \\
  Cont. $p$                          & $p_e/p_c$(\ion{C}{4})    &  0.105 & 0.696              & 16 \\
  Cont. $p$                          & $p_e/p_c$(\ion{C}{3}])    &  0.179 & 0.437              & 21 \\
  Cont. $p$                          & $p_e/p_c$(\ion{Mg}{2}) & -0.147 & 0.573              & 17 \\
  Cont. $p$                          & $p_a/p_c$(\ion{C}{4})    & -0.213 & 0.464              & 14 \\
  Cont. $p$                          & $p_a/p_c$(\ion{Mg}{2}) &  0.055 & 0.858              & 13 \\      
  $p_e/p_c$(\ion{C}{4})    & $p_e/p_c$(\ion{C}{3}])    &  0.553 & \textbf{0.026} & 16 \\                     
  $p_e/p_c$(\ion{C}{4})    & $p_e/p_c$(\ion{Mg}{2}) & -0.179 & 0.702               &  7 \\ 
  $p_e/p_c$(\ion{C}{4})    & $p_a/p_c$(\ion{C}{4})    &  0.476 & 0.118               & 12 \\
  $p_e/p_c$(\ion{C}{4})    & $p_a/p_c$(\ion{Mg}{2}) & -1.00   & \textbf{0.00}    & 3  \\
  $p_e/p_c$(\ion{C}{3}])    & $p_e/p_c$(\ion{Mg}{2}) &  0.103 & 0.777               & 10 \\
  $p_e/p_c$(\ion{C}{3}])    & $p_a/p_c$(\ion{C}{4})    &  0.478 & 0.099               & 13 \\
  $p_e/p_c$(\ion{C}{3}])    & $p_a/p_c$(\ion{Mg}{2}) &  0.657 & 0.156                &  6 \\
  $p_e/p_c$(\ion{Mg}{2}) & $p_a/p_c$(\ion{C}{4})    & -0.300 & 0.624               &  5 \\
  $p_e/p_c$(\ion{Mg}{2}) & $p_a/p_c$(\ion{Mg}{2}) &  0.797  & \textbf{0.003} & 11 \\
  $p_a/p_c$(\ion{C}{4})    & $p_a/p_c$(\ion{Mg}{2}) & -0.500 & 0.667               &  3 \\				       
 \enddata
 \tablecomments{The number of points considered in each correlation is given in the column labeled $n$.}
\end{deluxetable}

\clearpage

\begin{deluxetable}{cccccccc}
 \tabletypesize{\scriptsize}
 \tablewidth{0pt}
 \tablecaption{Other radio-selected BALQSO spectropolarimetry from the literature.\label{othertbl}}
 \tablehead{
   \colhead{Object} & \colhead{z} & \colhead{logR*} & \colhead{P(\%)} & \colhead{PA($^{\circ}$)} & \colhead{Wavelengths (\AA)} & \colhead{Class} & \colhead{Reference}
   }
  \startdata
  FIRST J0840+3633 & 1.23 & 0.45                 & 2.74$\pm$0.06 & 47.4$\pm$0.7               & 4800-5100 & FeLoBAL & Brotherton et al. (1997) \\
  FIRST J1556+3517 & 1.49 &  \nodata                  & 6.80$\pm$0.20 & 152.7$\pm$0.9              & 6000-6300 & FeLoBAL & Brotherton et al. (1997) \\
  QSO 2359-1241    & 0.87 & 1.66                 & $\sim$4-6     & 140-150                    & 4000-8600 & LoBAL   & Brotherton et al. (2001) \\
  FIRST J1016+5209 & 2.46 & 2.7\tablenotemark{a} & $\sim$2.5     & 75-85                      & 4200-8000 & HiBAL   & Gregg et al. (2000)\\
  LBQS 1138-0126   & 1.27 & 2.5                  & $\sim$3.5     & $\sim$0                    & 3850-4100 & HiBAL   & Brotherton et al. (2002) \\
  PKS 0040-005     & 2.81 & 2.7                  & 0.72$\pm$0.08 & 124$\pm$4\tablenotemark{b} & 5000-6600 & HiBAL & Brotherton et al. (2006) \\
  \enddata
  \tablecomments{Measurements are continuum or broadband over the wavelengths indicated}
 \tablenotetext{a}{If only the core radio flux is considered, logR* becomes 2.0}
 \tablenotetext{b}{The polarization position angle published in Brotherton et al. (2006) has been discovered to be incorrect due to a reduction error; an erratum is being prepared.}
\end{deluxetable}

\clearpage

\begin{figure}
 \centering
  \figurenum{2}
   \subfloat[][Fig. 2$a$]{\includegraphics[width=5in]{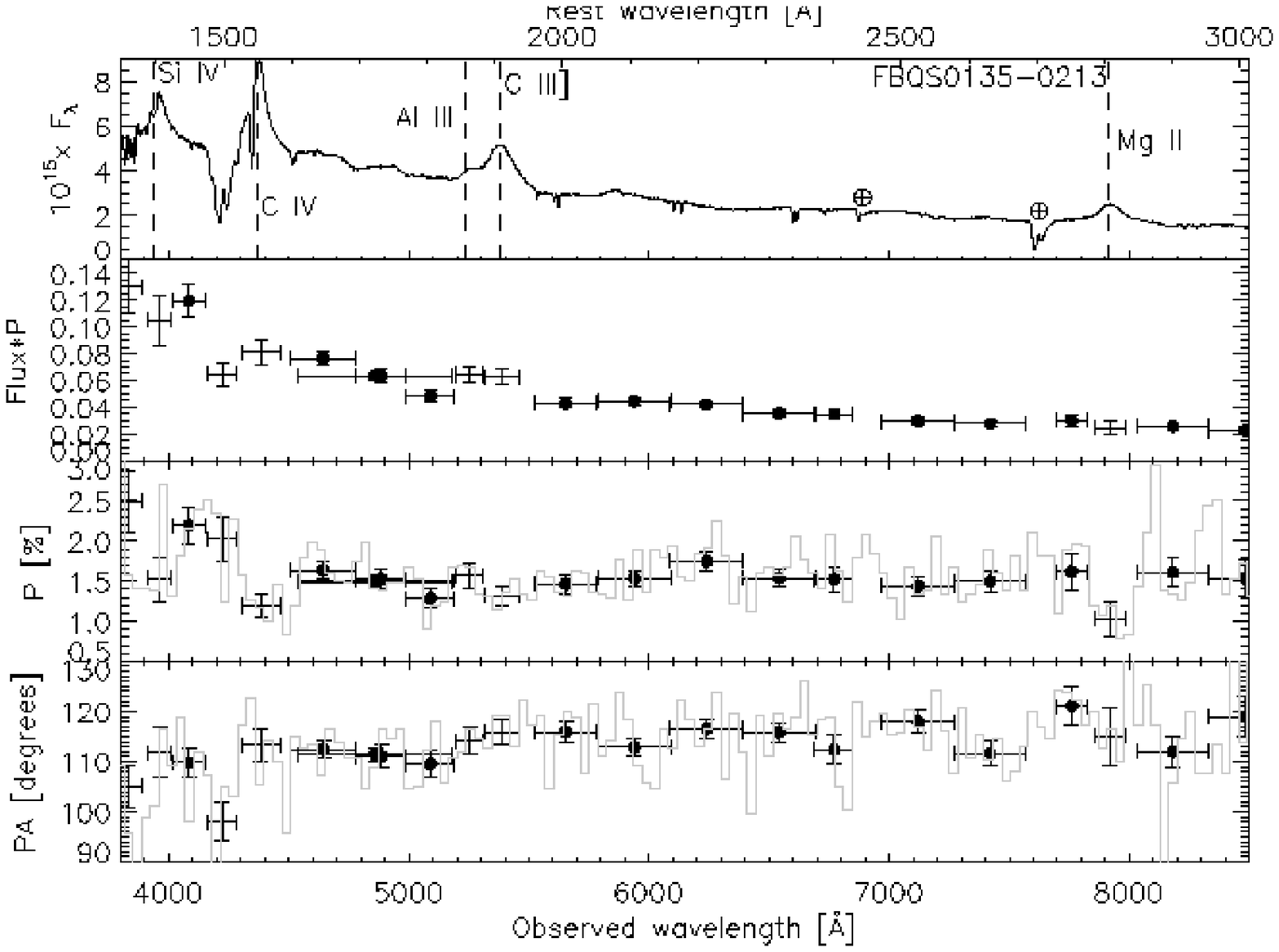}}
  \caption{The following figures show the spectropolarimetry for our whole sample.  The top panel is the total flux spectrum, followed by the polarized flux spectrum (P times f), the polarization level, and finally the polarization position angle.  Error bars are 1-$\sigma$ values. The light colored lines in the polarization and position angle plots are smaller equal sized bins (unbiased, with errors omitted) which range from 15-35 \AA\ wide depending on the signal-to-noise of the observations.  Observed frame wavelengths are given on the bottom, and rest frame wavelengths are given along the top.  The name of each object is in the total flux panel.  Prominent lines are marked in this first figure as an aid to the reader; lines marked with $\oplus$ are telluric lines.  Bins plotted in bold in the polarization panel indicate where continuum measurements were made.  Points plotted with a closed circle are continuum measurements, those with just a point are line measurements.\label{polplotsfig}}
\end{figure}

\begin{figure}
 \ContinuedFloat
 \centering
  \subfloat[][Fig. 2$b$]{\includegraphics[width=5in]{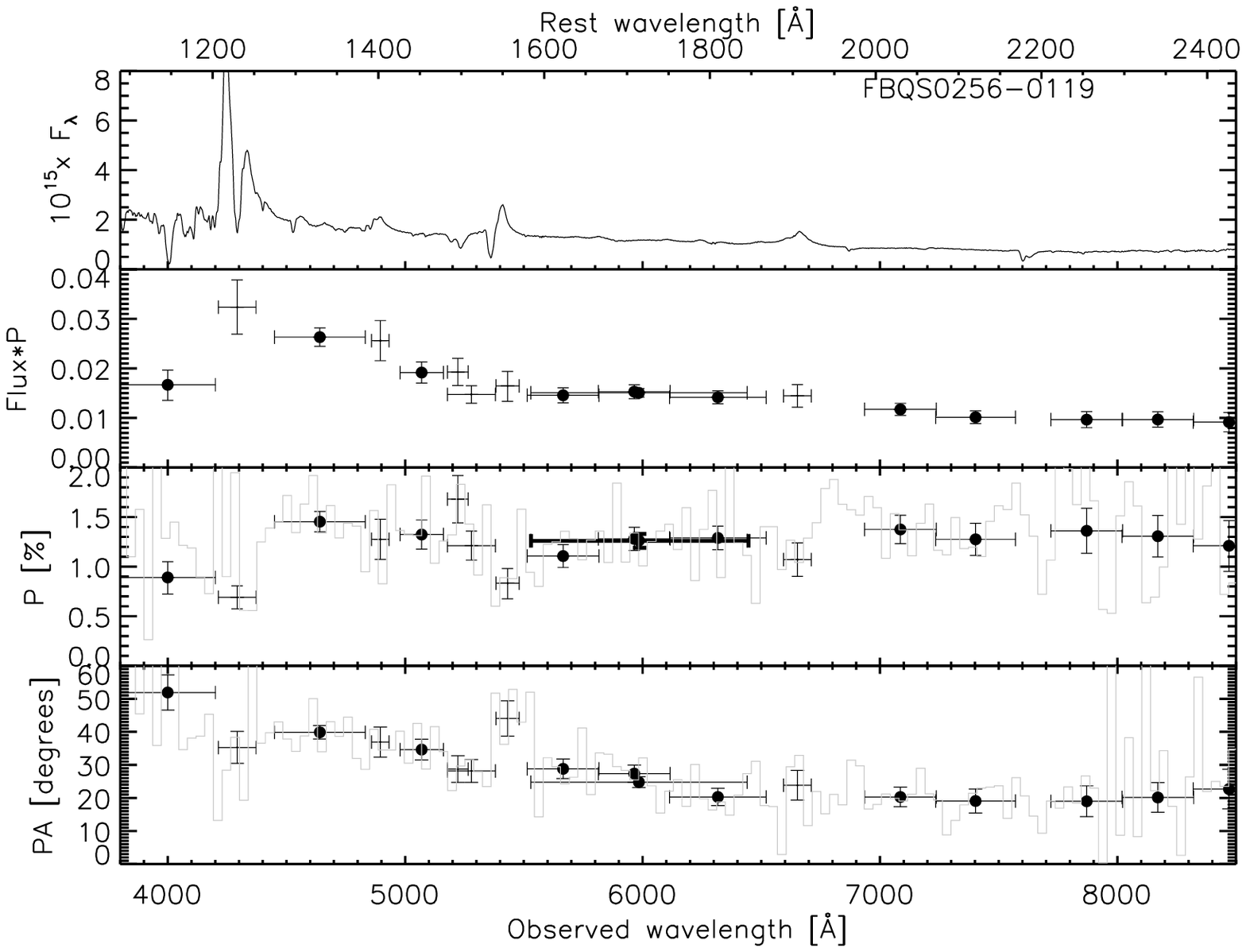}}
\end{figure}

\begin{figure}
 \ContinuedFloat
 \centering
  \subfloat[][Fig. 2$c$]{\includegraphics[width=5in]{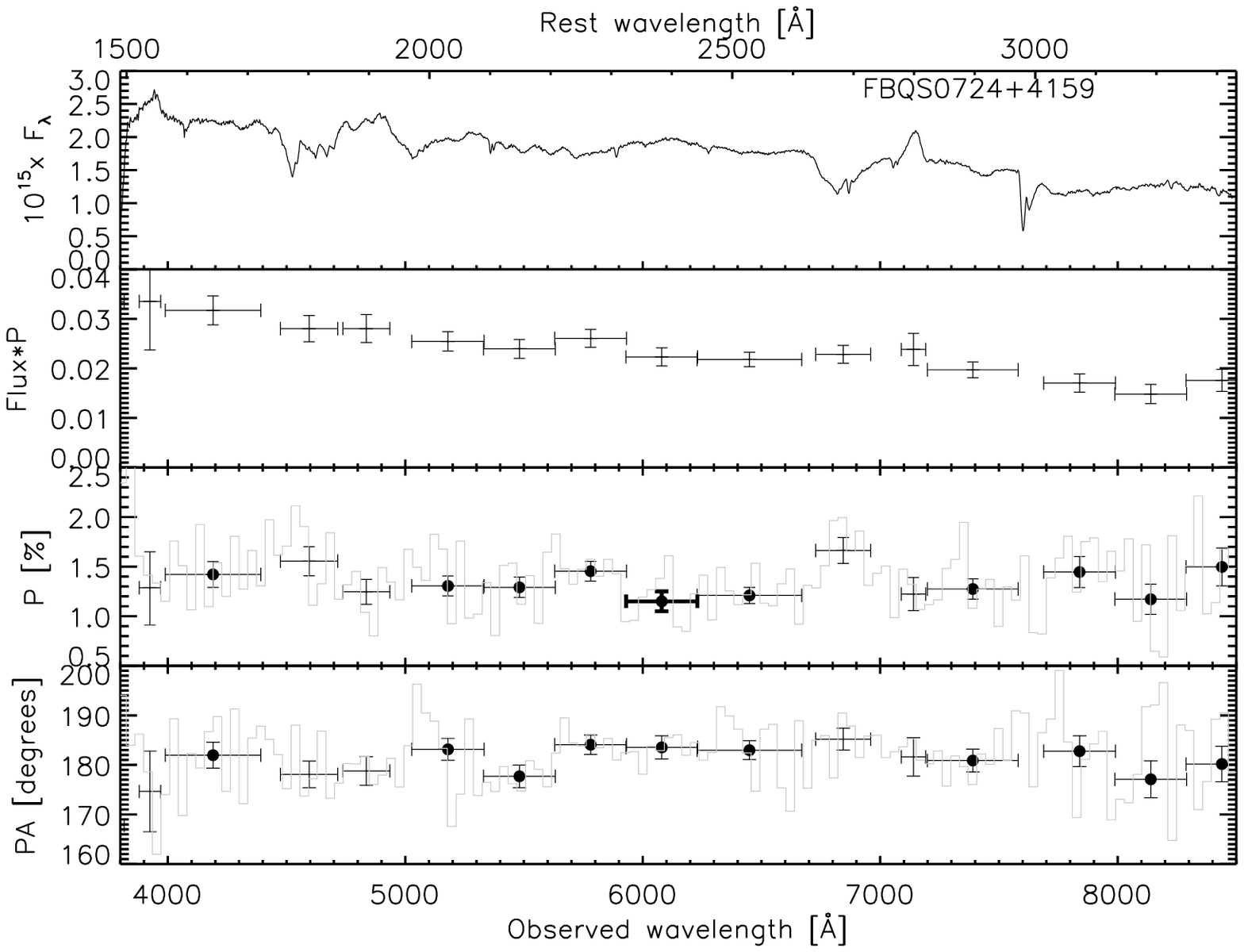}}
\end{figure}

\begin{figure}
 \ContinuedFloat
 \centering
  \subfloat[][Fig. 2$d$]{\includegraphics[width=5in]{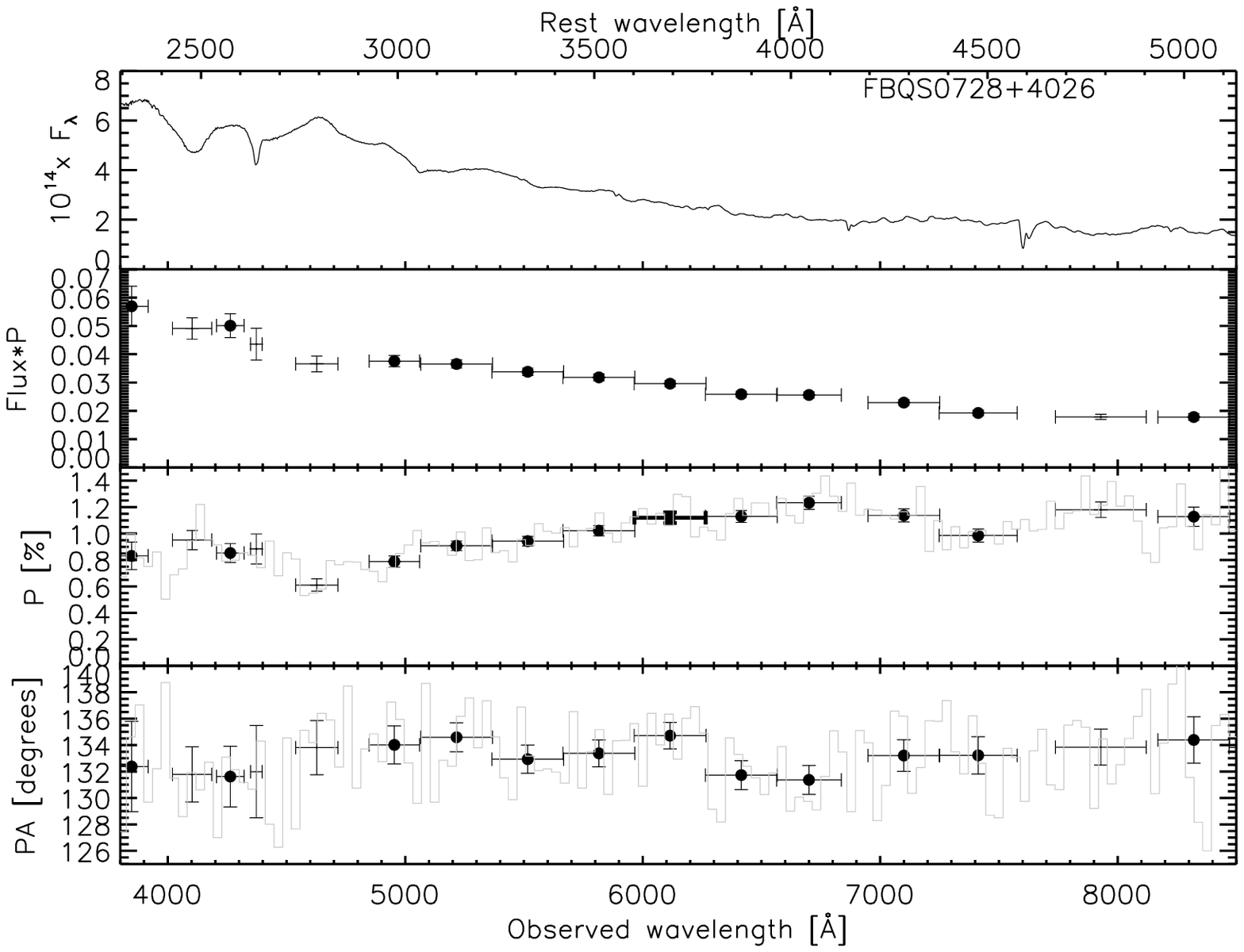}}
\end{figure}

\begin{figure}
 \ContinuedFloat
 \centering
  \subfloat[][Fig. 2$e$]{\includegraphics[width=5in]{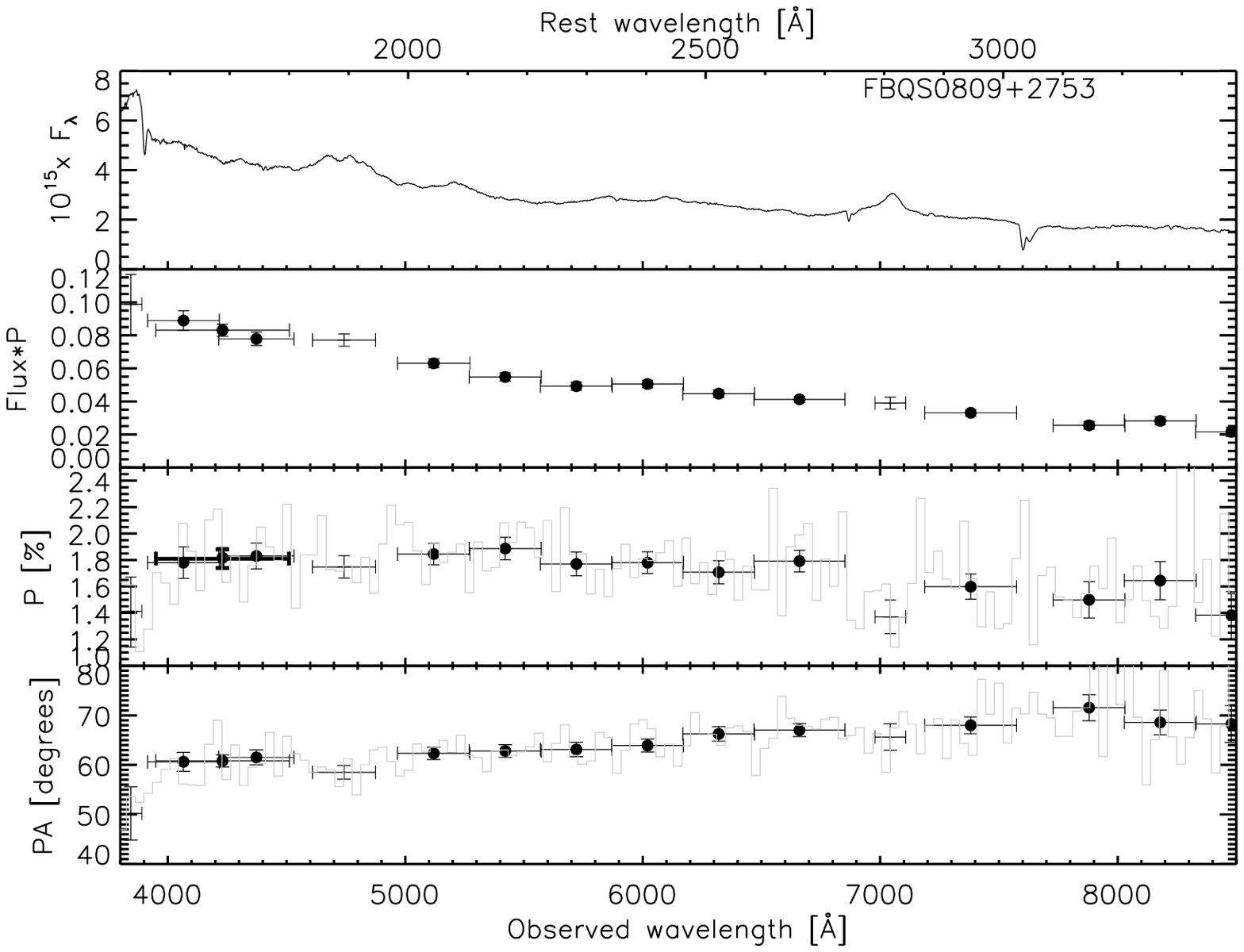}}
\end{figure}

\begin{figure}
 \ContinuedFloat
 \centering
  \subfloat[][Fig. 2$f$]{\includegraphics[width=5in]{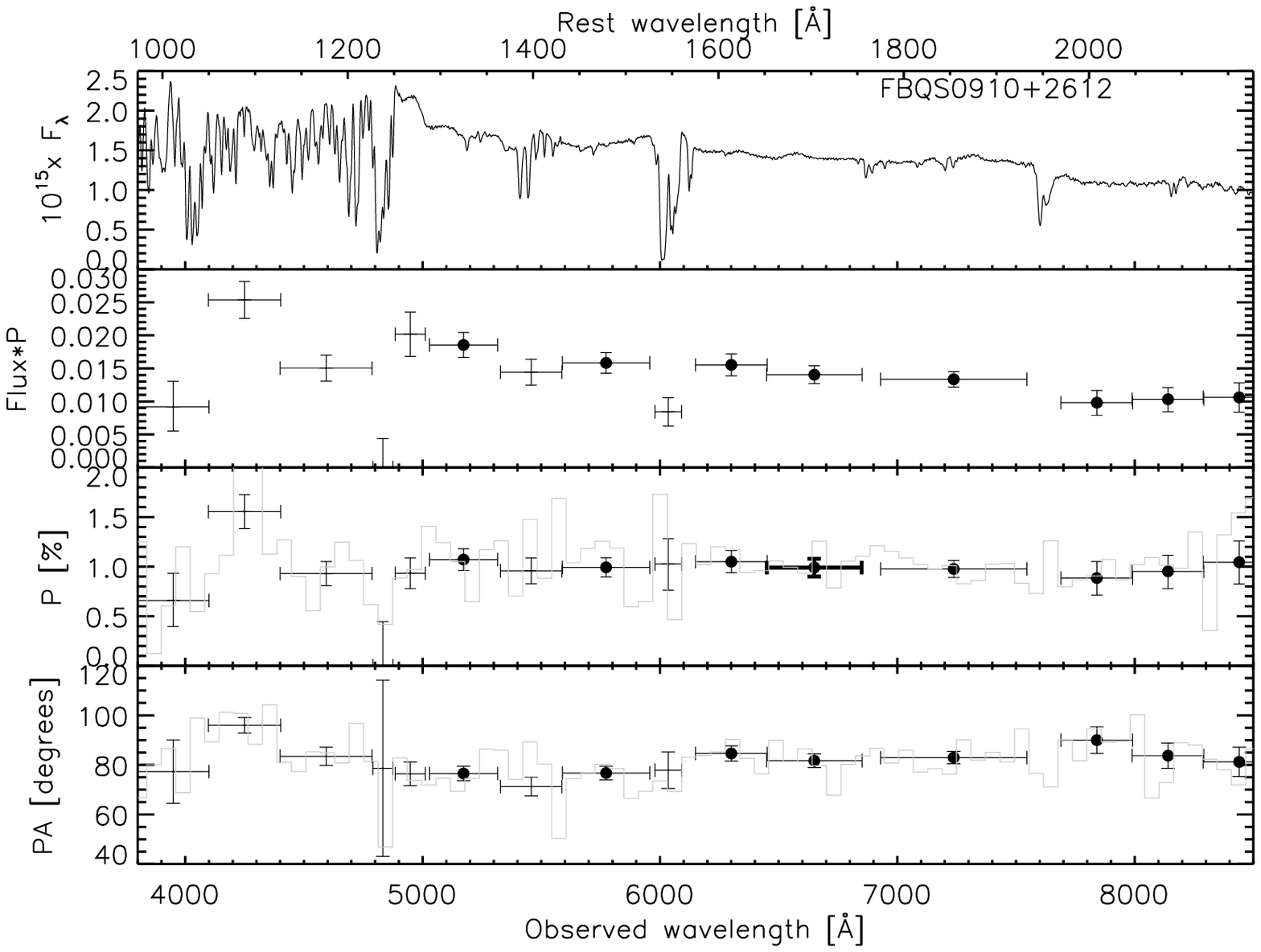}}
\end{figure}

\begin{figure}
 \ContinuedFloat
 \centering
  \subfloat[][Fig. 2$g$]{\includegraphics[width=5in]{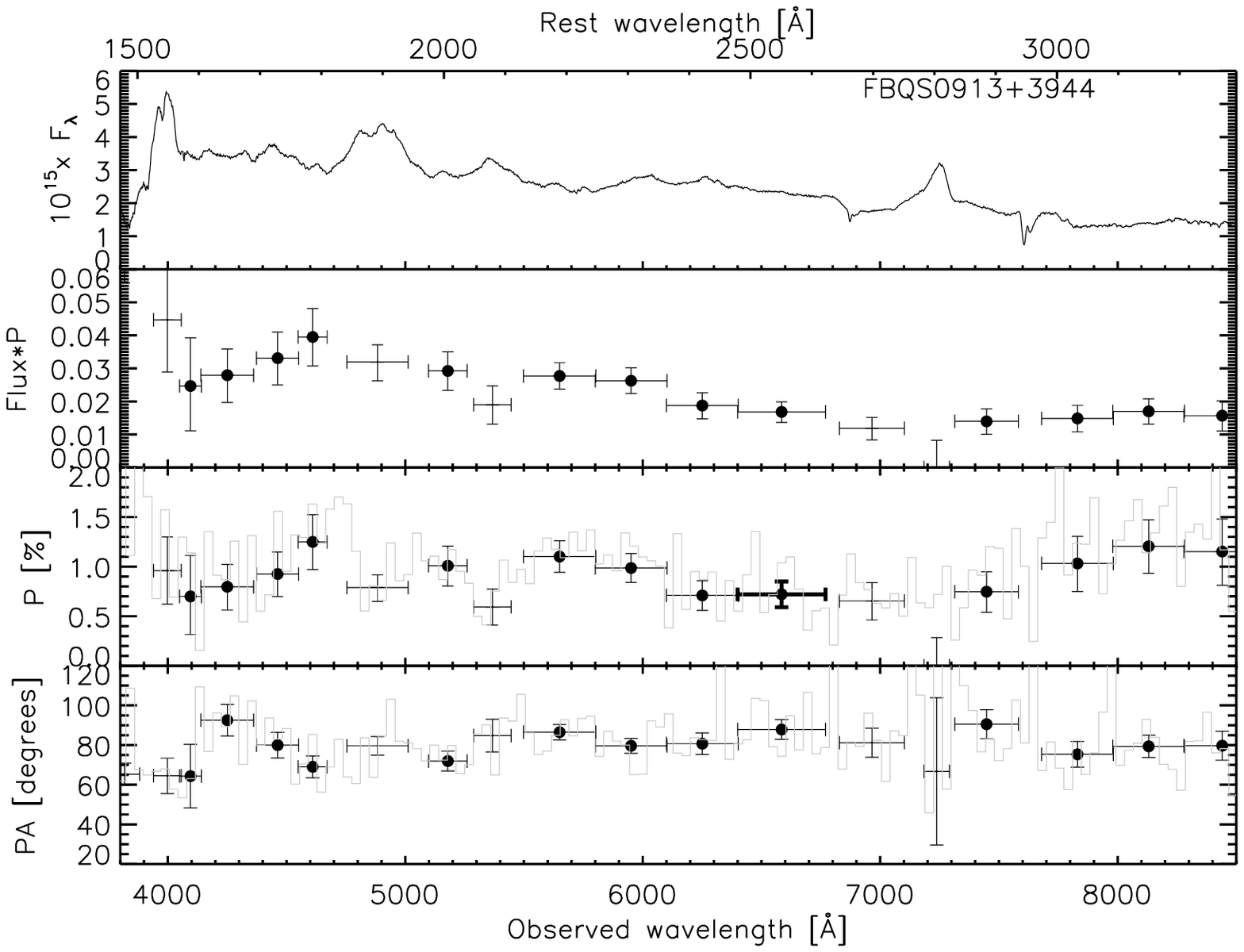}}
\end{figure}

\begin{figure}
 \ContinuedFloat
 \centering
  \subfloat[][Fig. 2$h$]{\includegraphics[width=5in]{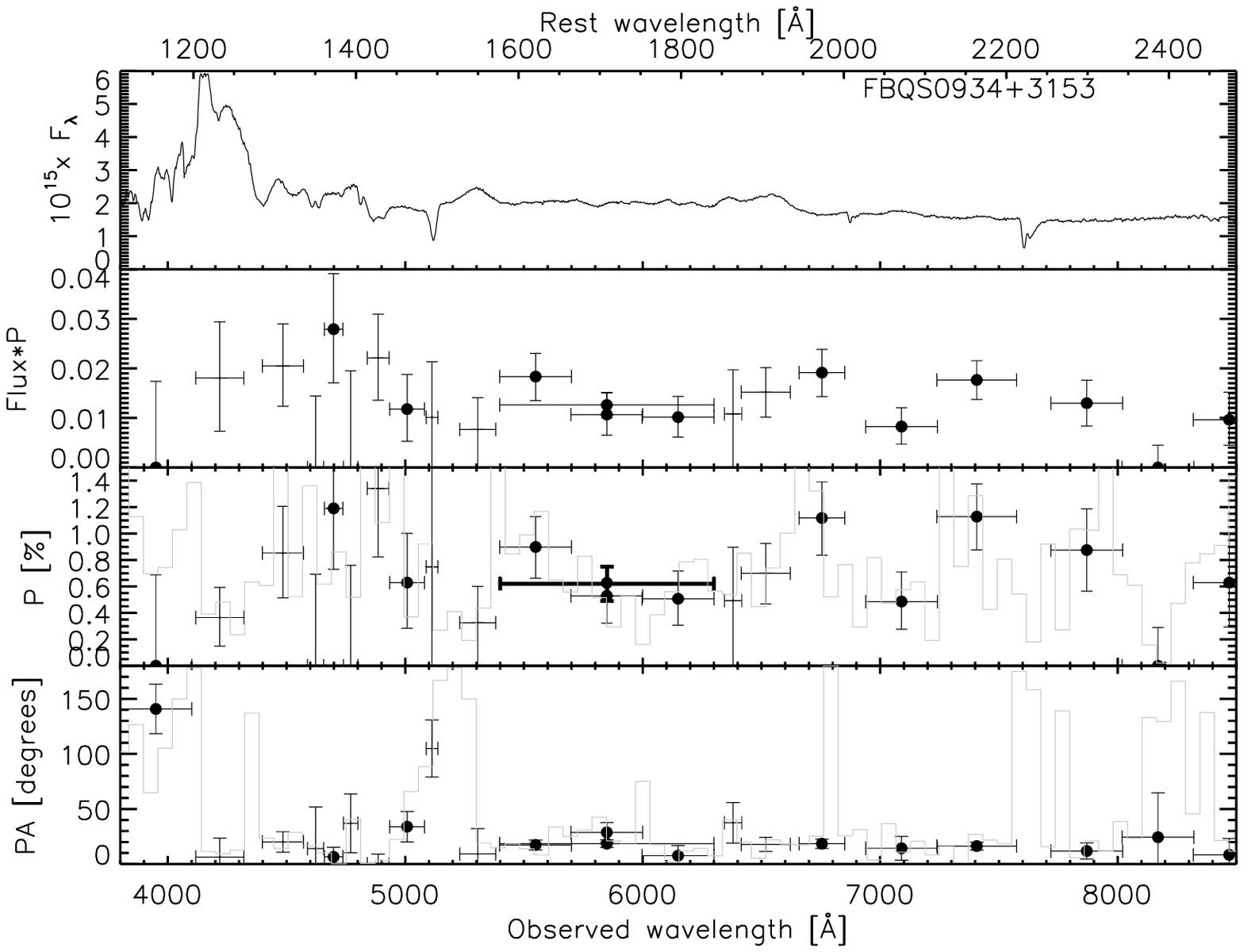}}
\end{figure}

\begin{figure}
 \ContinuedFloat
 \centering
  \subfloat[][Fig. 2$i$]{\includegraphics[width=5in]{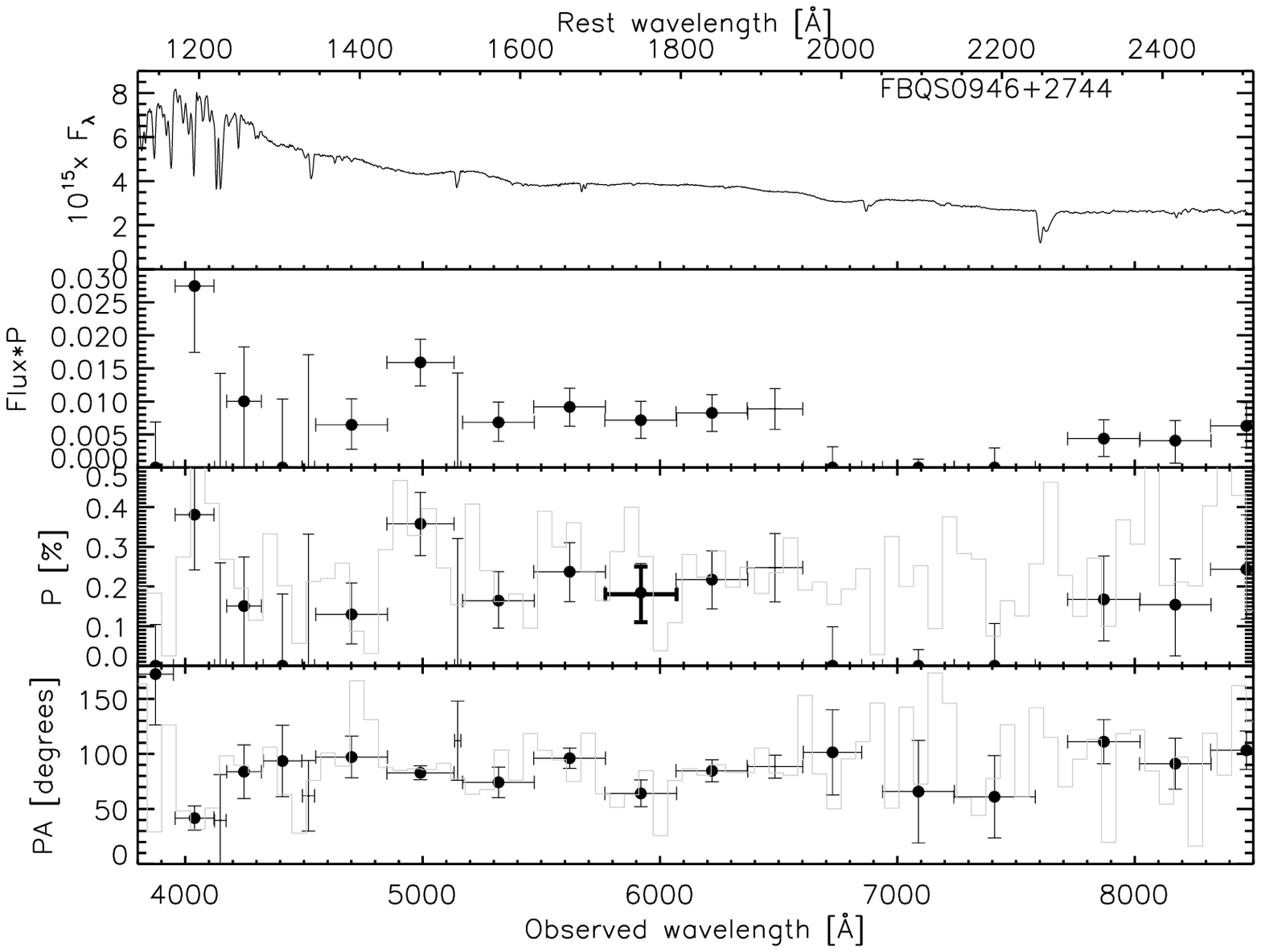}}
\end{figure}

\begin{figure}
 \ContinuedFloat
 \centering
  \subfloat[][Fig. 2$j$]{\includegraphics[width=5in]{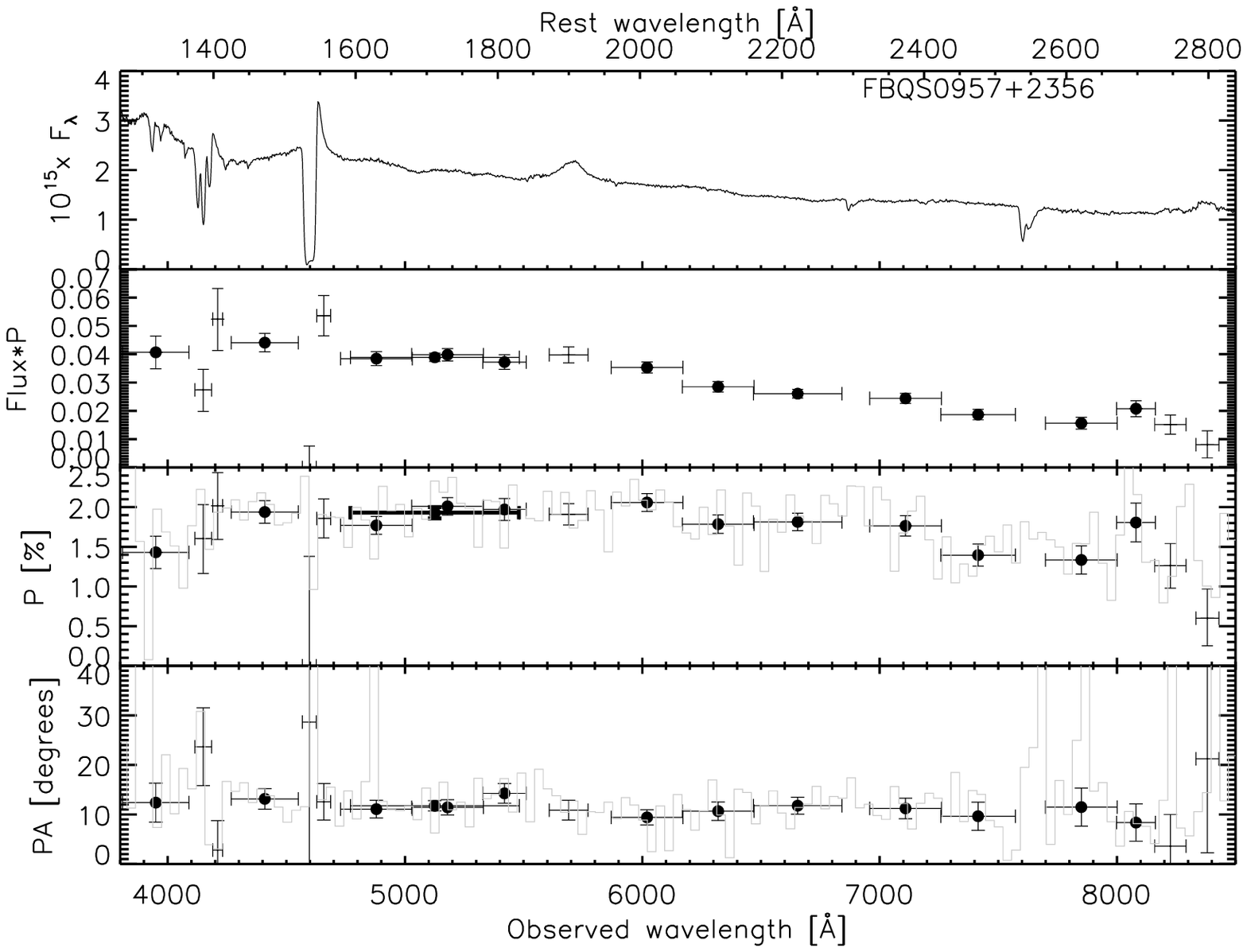}}
\end{figure}

\begin{figure}
 \ContinuedFloat
 \centering
  \subfloat[][Fig. 2$k$]{\includegraphics[width=5in]{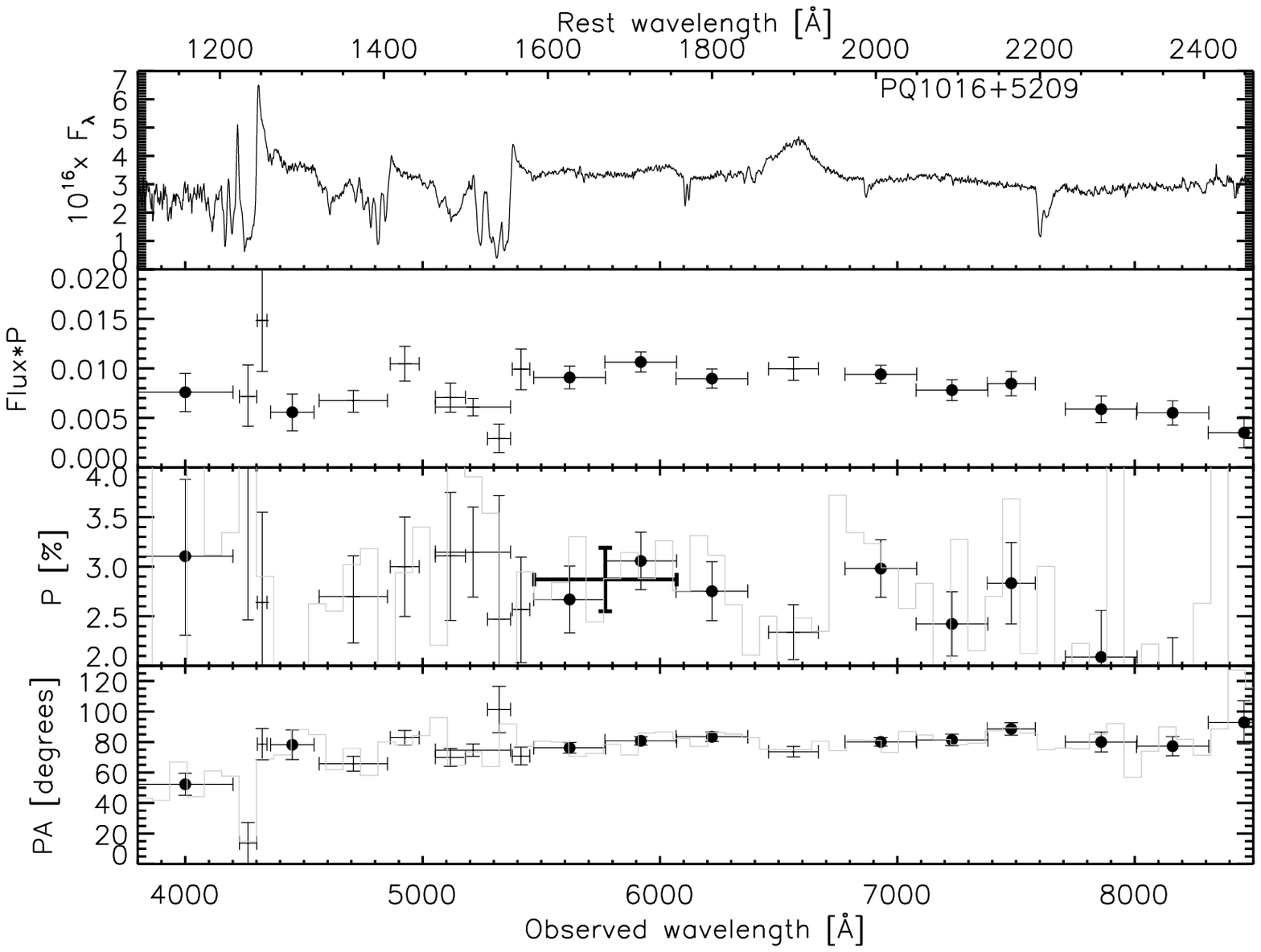}}
\end{figure}

\begin{figure}
 \ContinuedFloat
 \centering
  \subfloat[][Fig. 2$l$]{\includegraphics[width=5in]{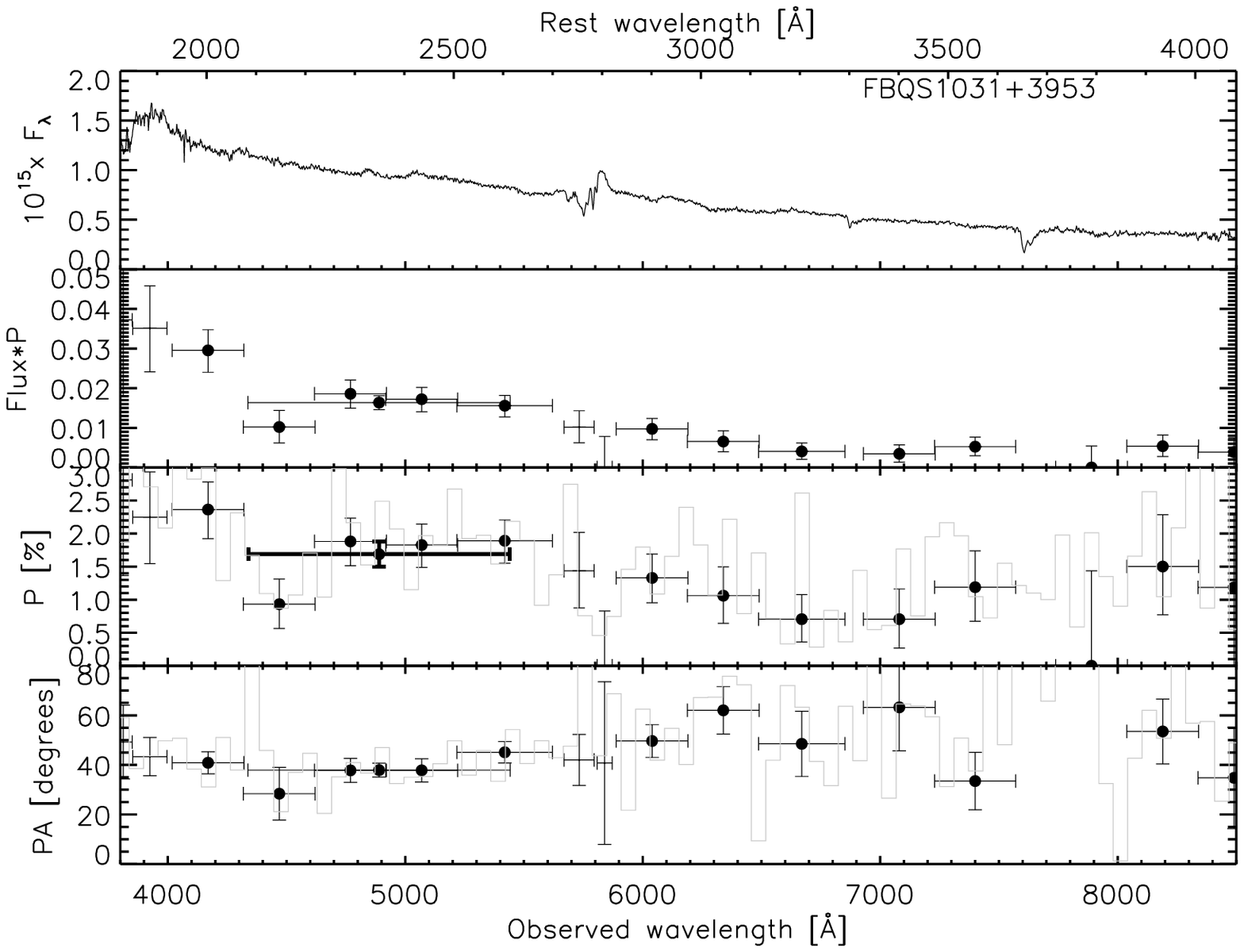}}
\end{figure}

\begin{figure}
 \ContinuedFloat
 \centering
  \subfloat[][Fig. 2$m$]{\includegraphics[width=5in]{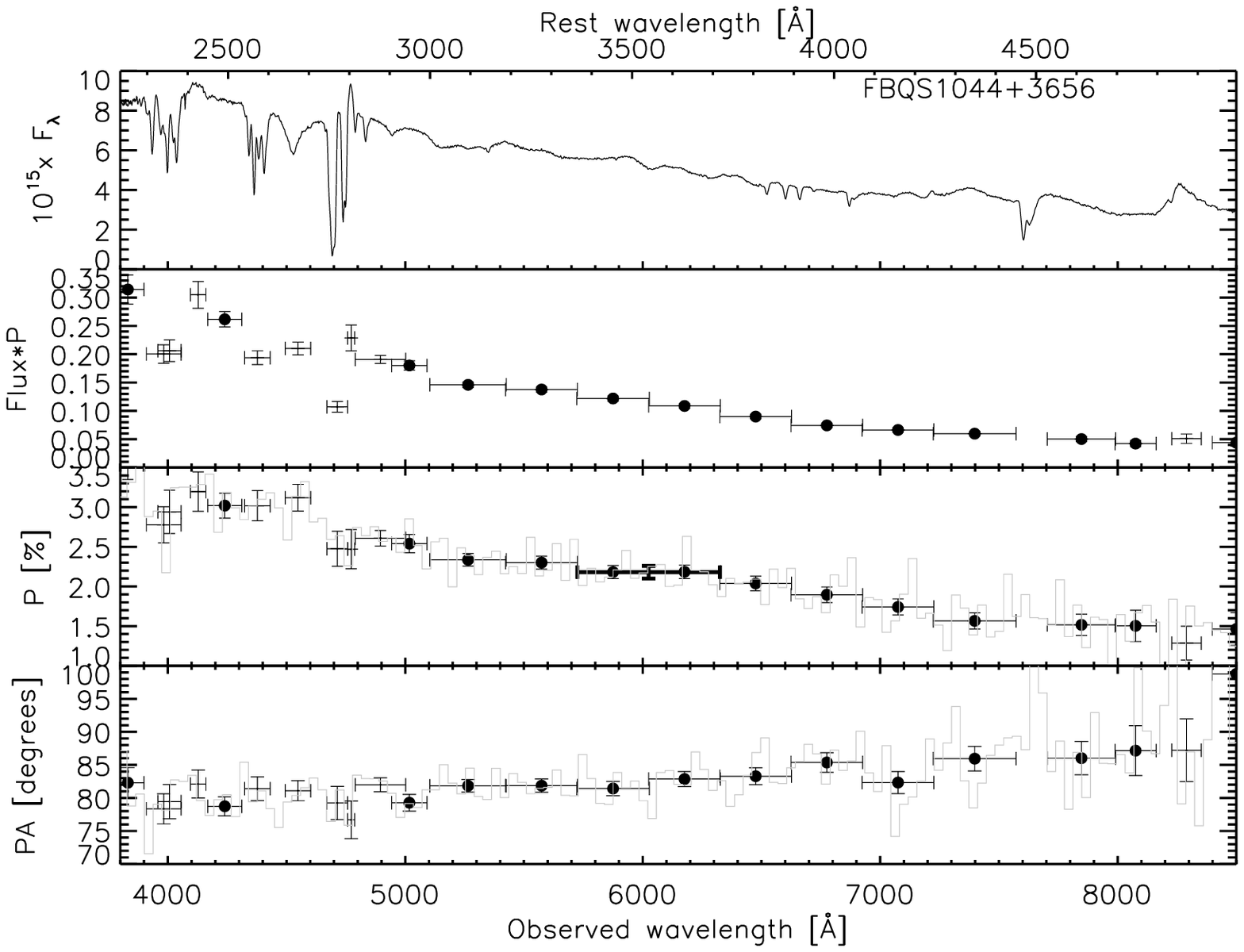}}
\end{figure}

\begin{figure}
 \ContinuedFloat
 \centering
  \subfloat[][Fig. 2$n$]{\includegraphics[width=5in]{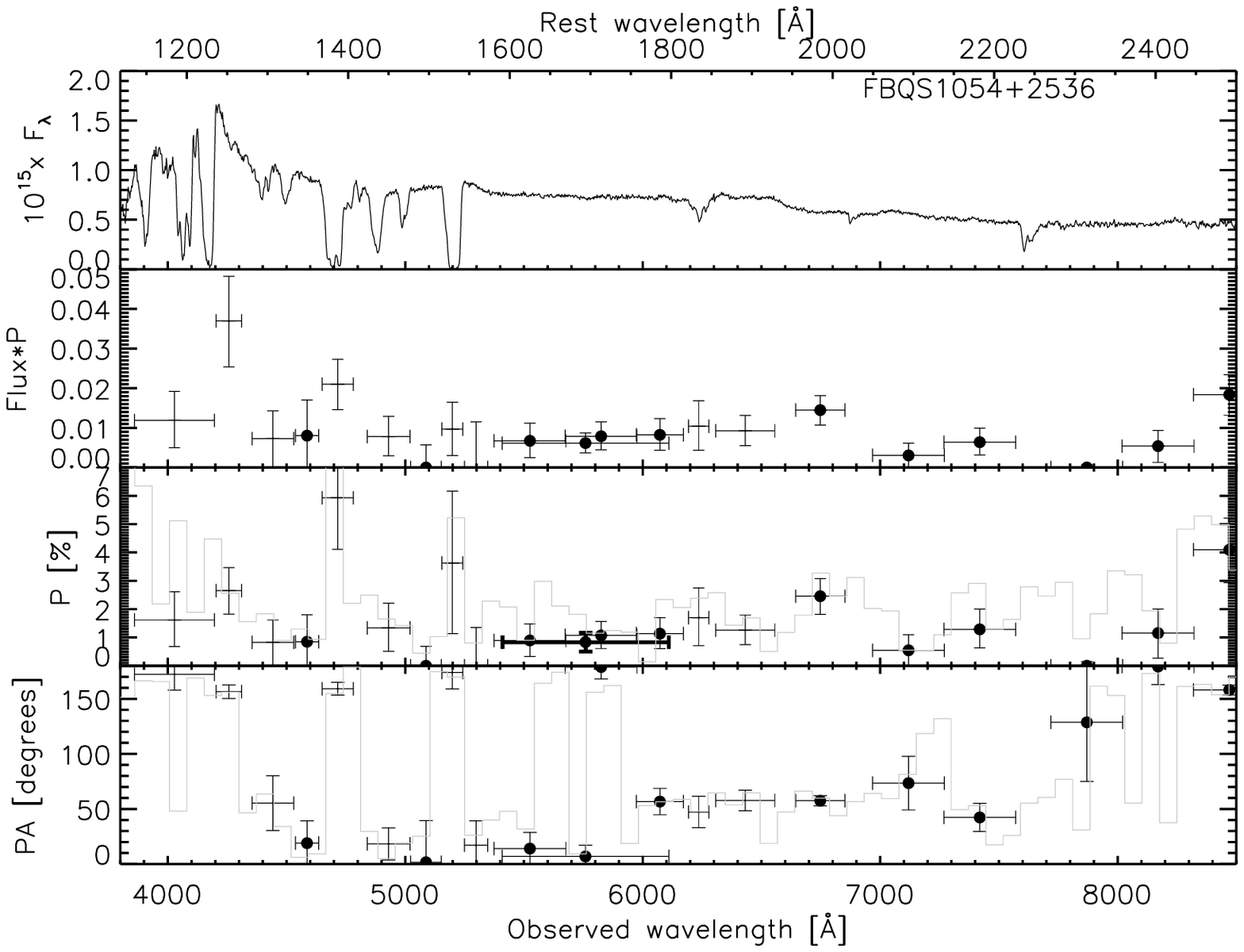}}
\end{figure}

\begin{figure}
 \ContinuedFloat
 \centering
  \subfloat[][Fig. 2$o$]{\includegraphics[width=5in]{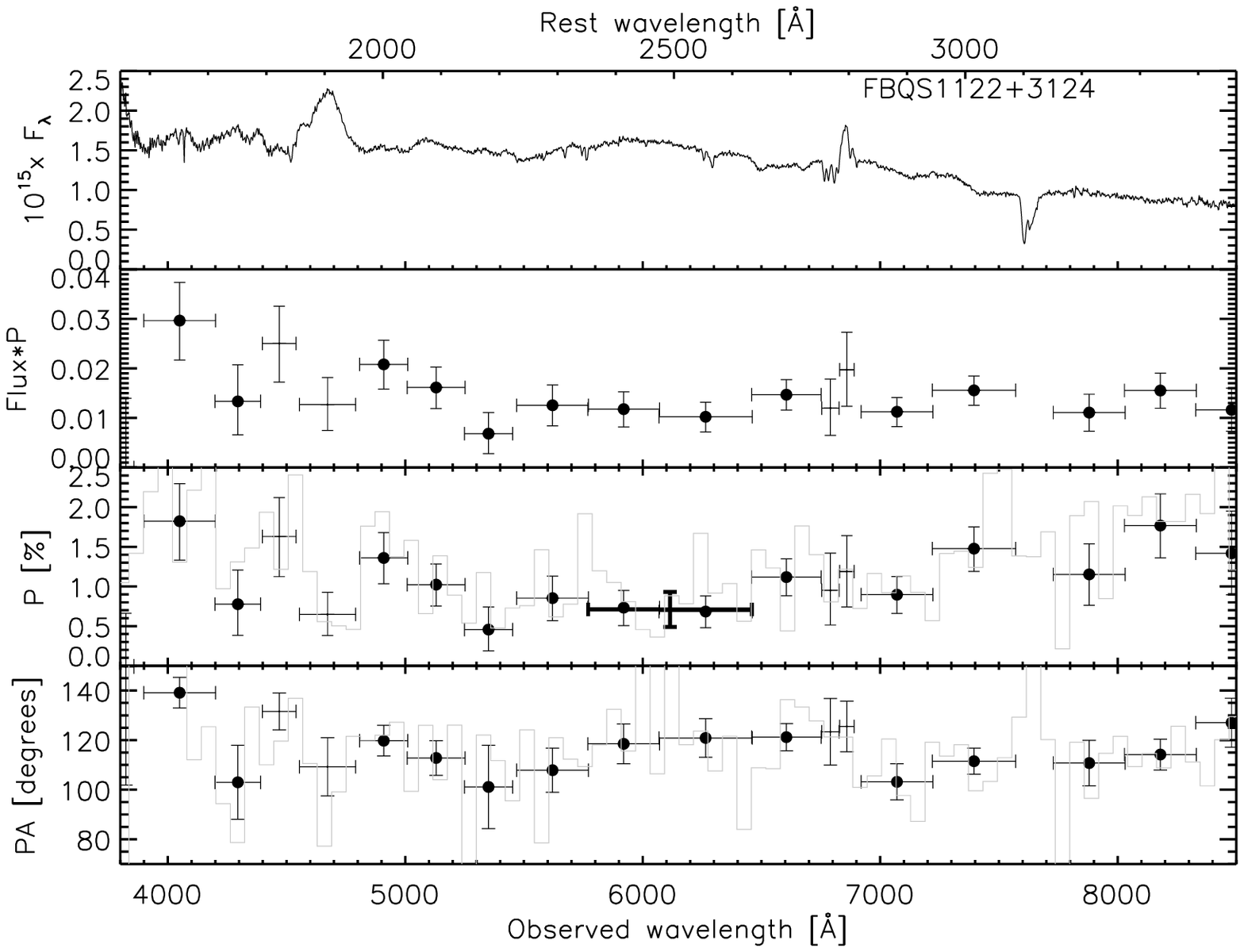}}
\end{figure}

\begin{figure}
 \ContinuedFloat
 \centering
  \subfloat[][Fig. 2$p$]{\includegraphics[width=5in]{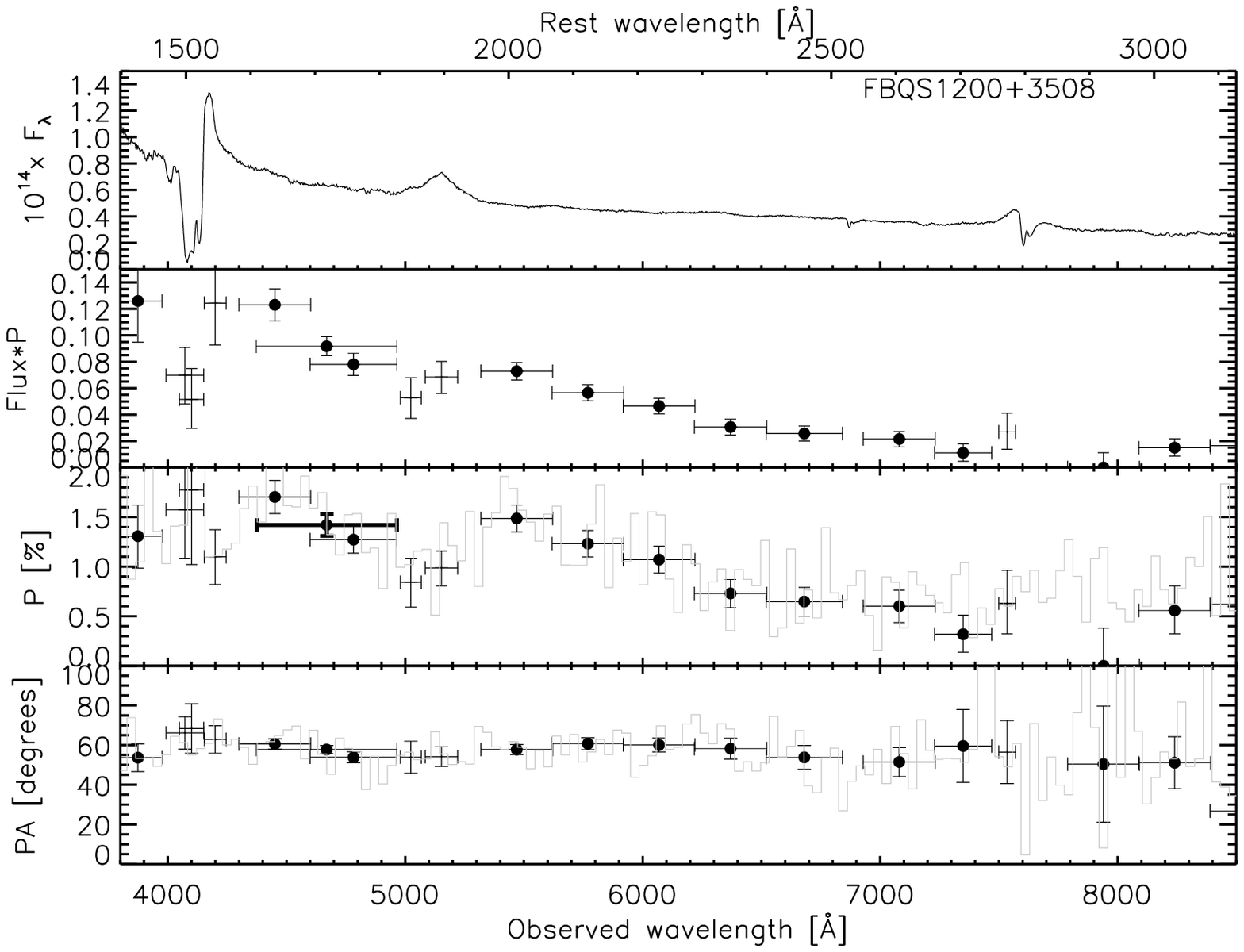}}
\end{figure}

\begin{figure}
 \ContinuedFloat
 \centering
  \subfloat[][Fig. 2$q$]{\includegraphics[width=5in]{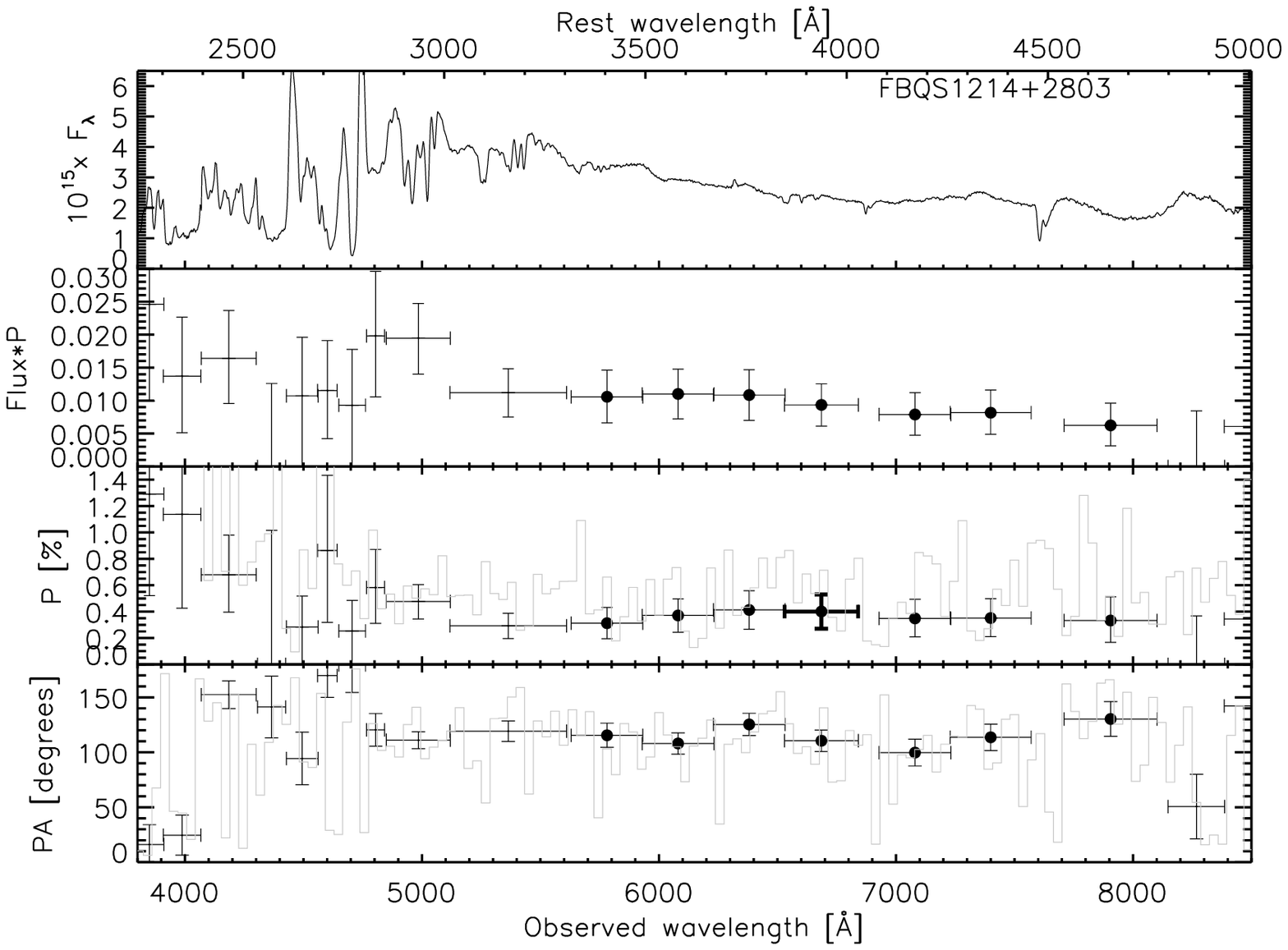}}
\end{figure}

\clearpage

\begin{figure}
 \ContinuedFloat
 \centering
  \subfloat[][Fig. 2$r$]{\includegraphics[width=5in]{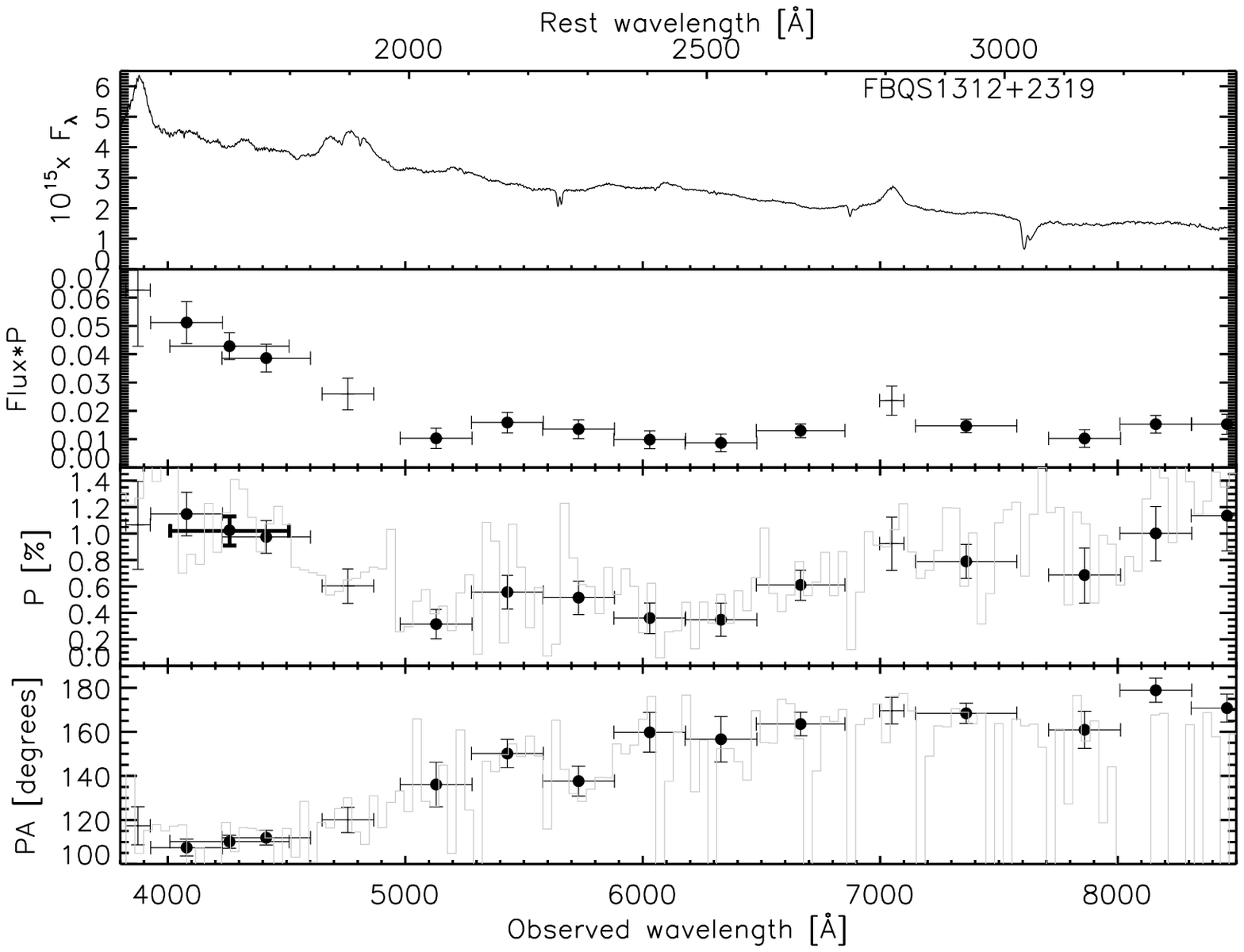}}
\end{figure}

\begin{figure}
 \ContinuedFloat
 \centering
  \subfloat[][Fig. 2$s$]{\includegraphics[width=5in]{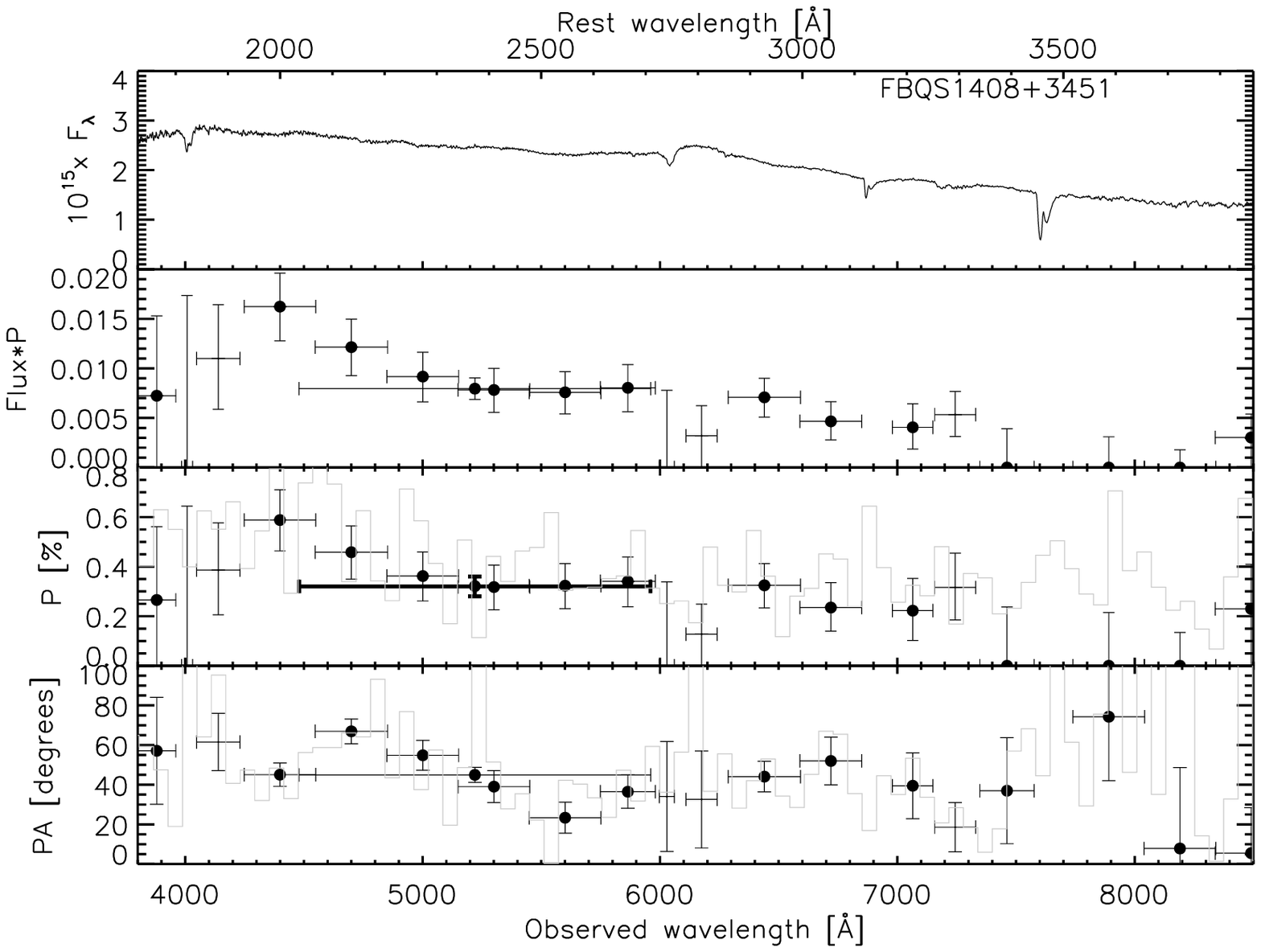}}
\end{figure}

\begin{figure}
 \ContinuedFloat
 \centering
  \subfloat[][Fig. 2$t$]{\includegraphics[width=5in]{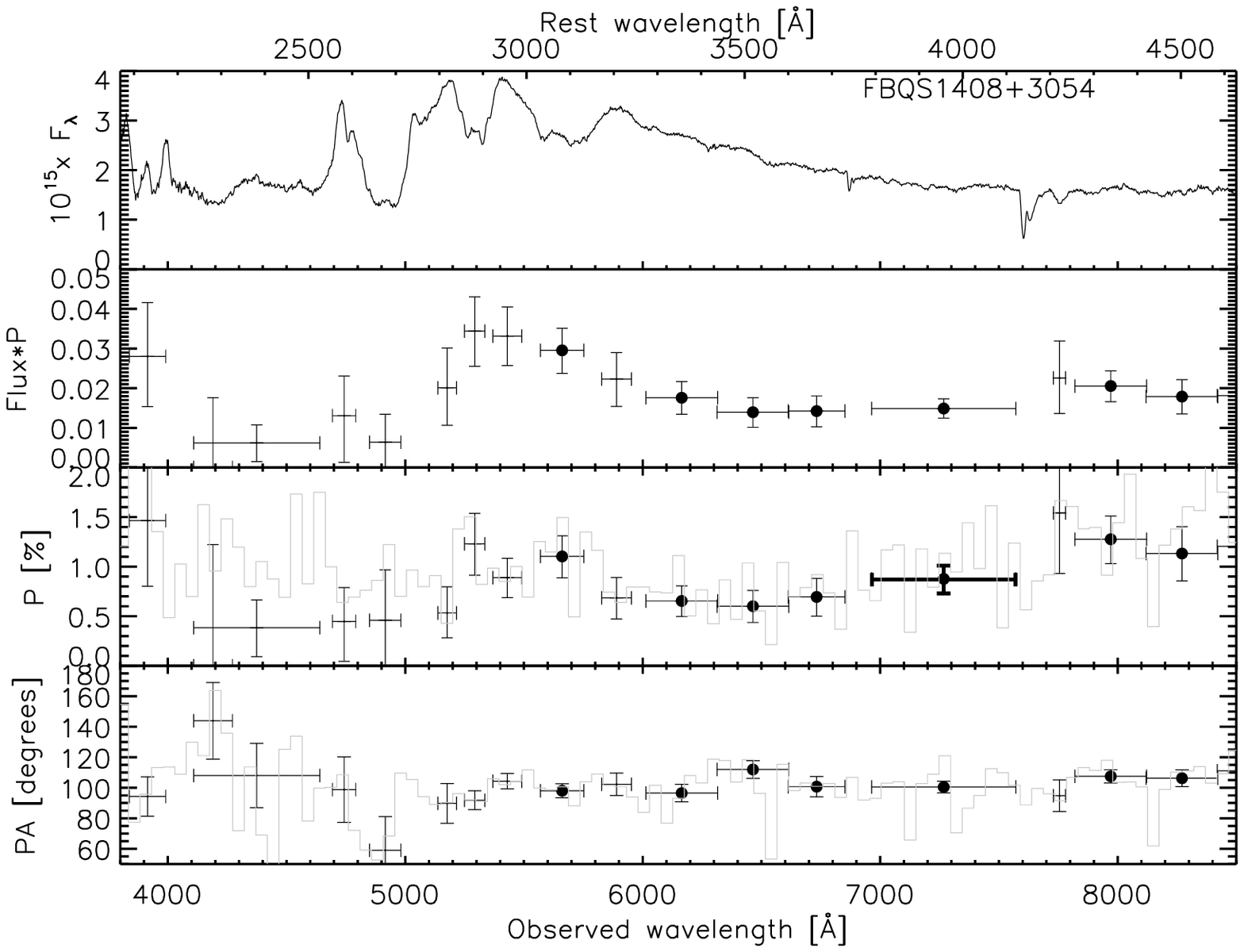}}
\end{figure}

\begin{figure}
 \ContinuedFloat
 \centering
  \subfloat[][Fig. 2$u$]{\includegraphics[width=5in]{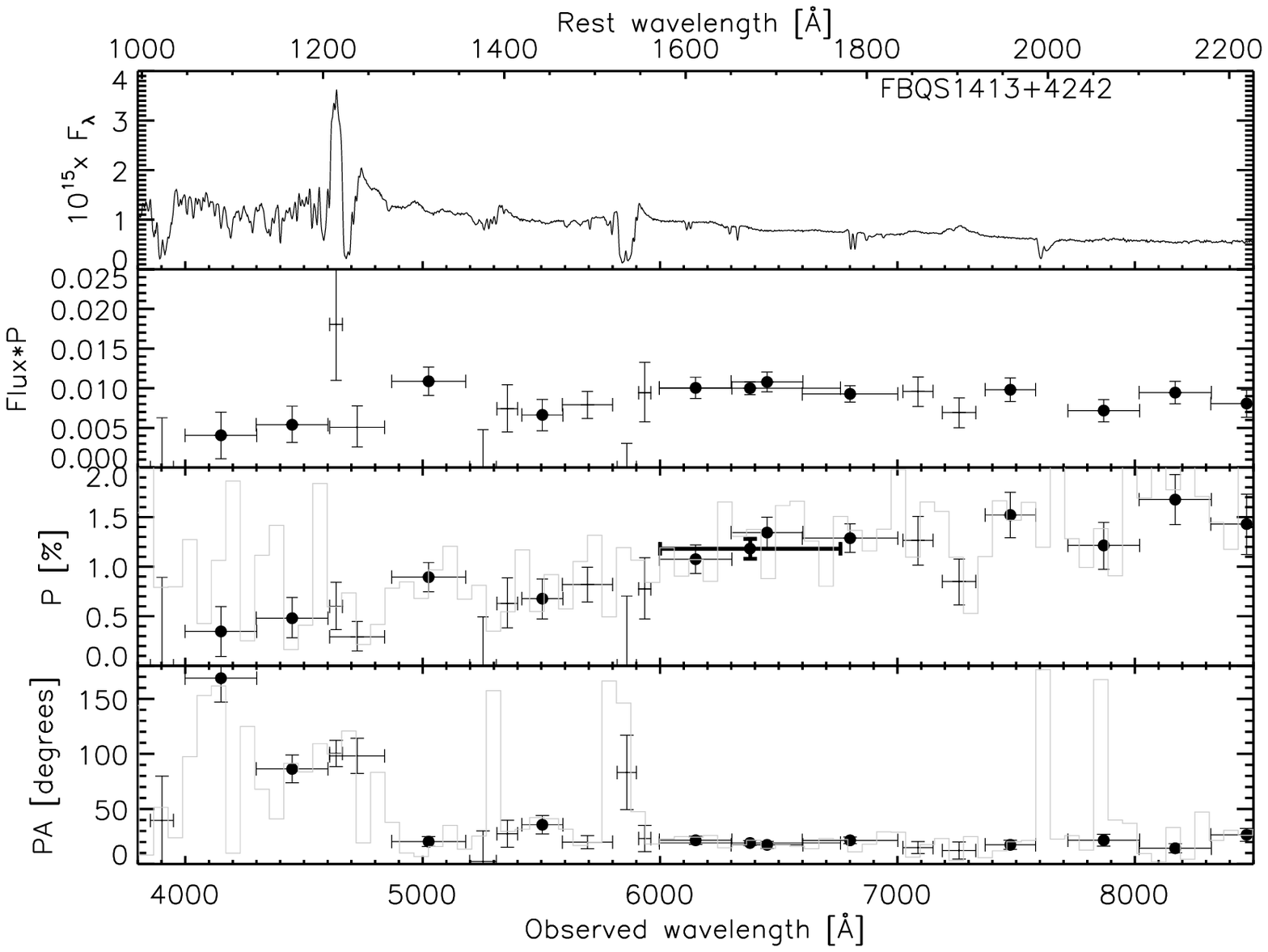}}
\end{figure}

\begin{figure}
 \ContinuedFloat
 \centering
  \subfloat[][Fig. 2$v$]{\includegraphics[width=5in]{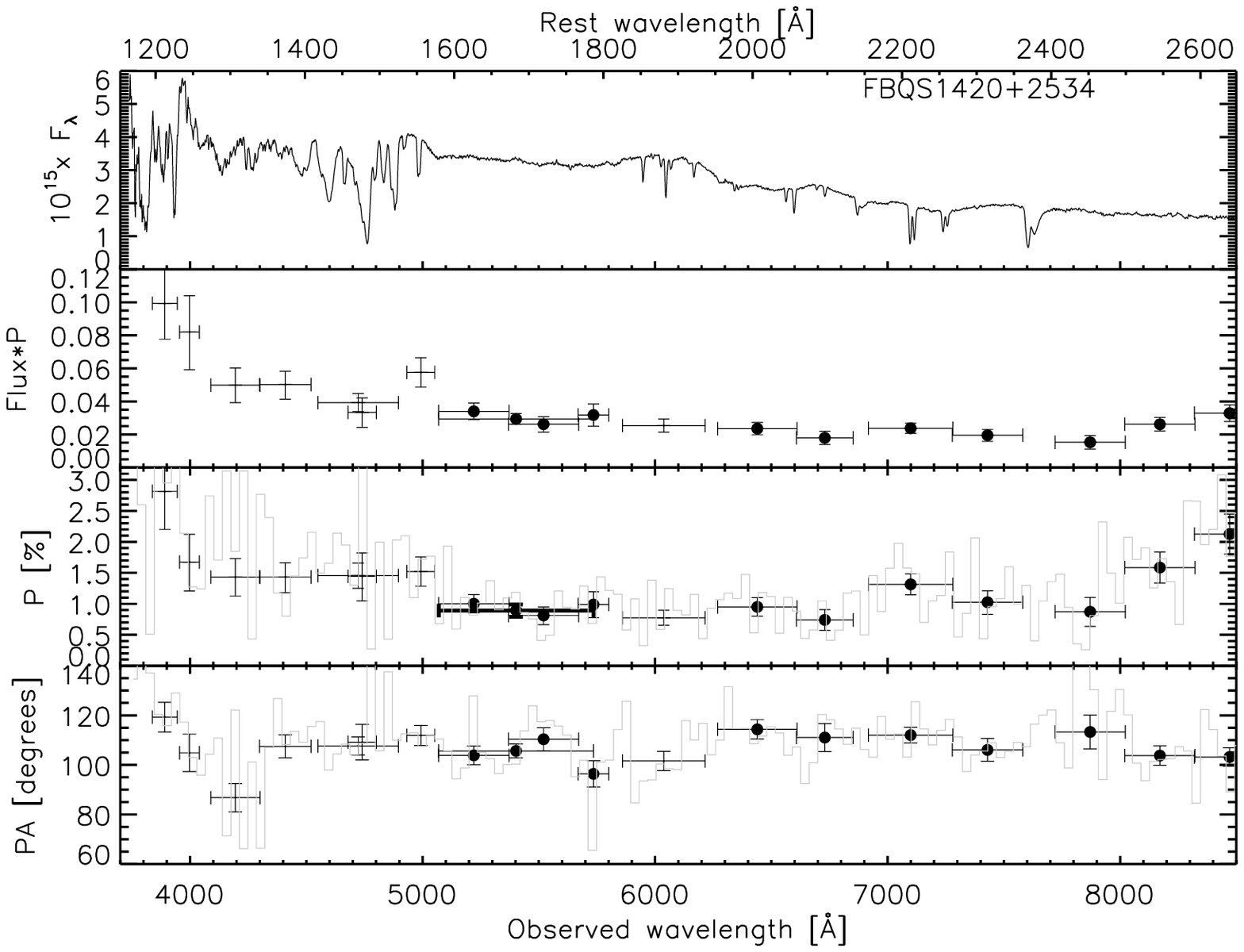}}
\end{figure}

\begin{figure}
 \ContinuedFloat
 \centering
  \subfloat[][Fig. 2$w$]{\includegraphics[width=5in]{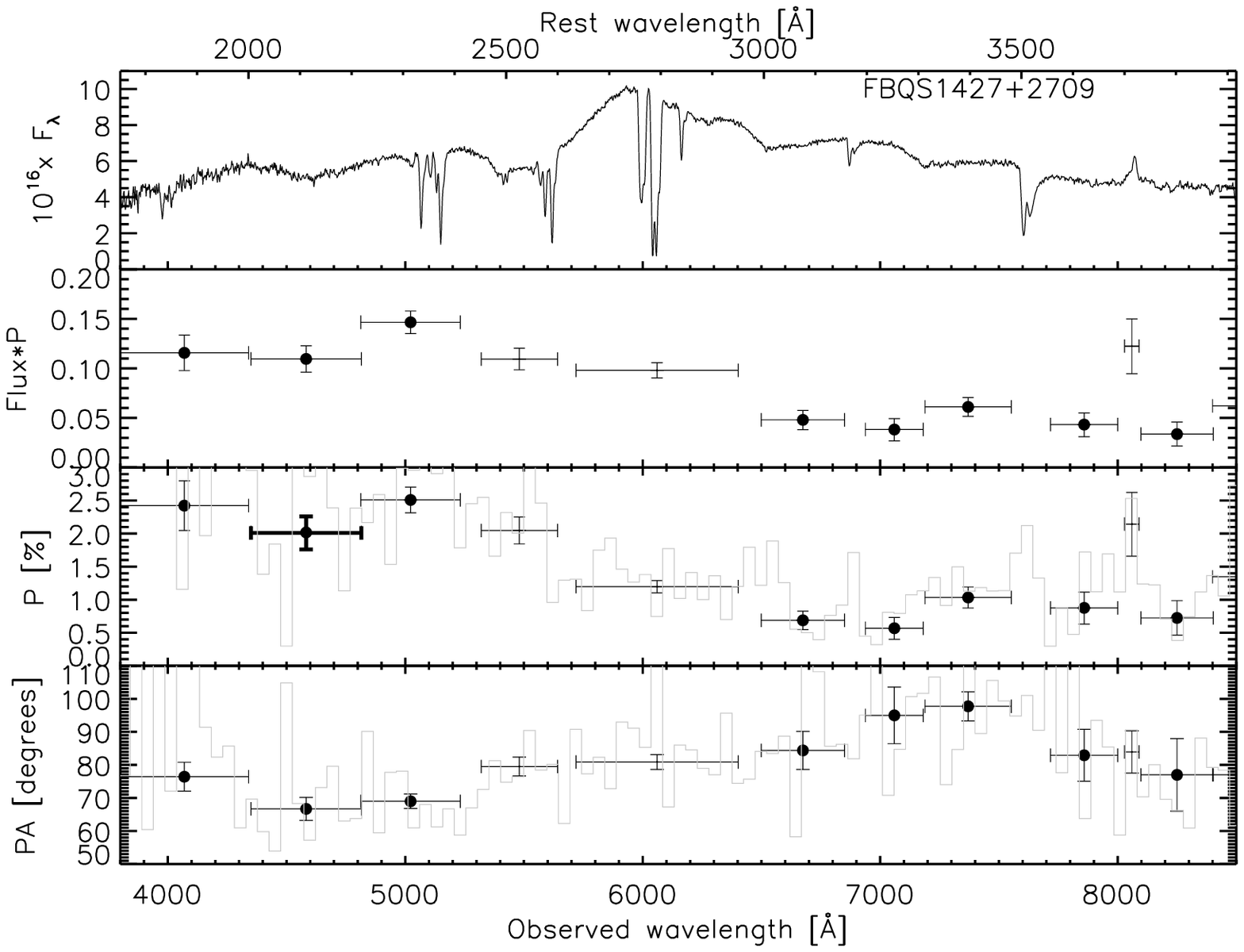}}
\end{figure}

\begin{figure}
 \ContinuedFloat
 \centering
  \subfloat[][Fig. 2$x$]{\includegraphics[width=5in]{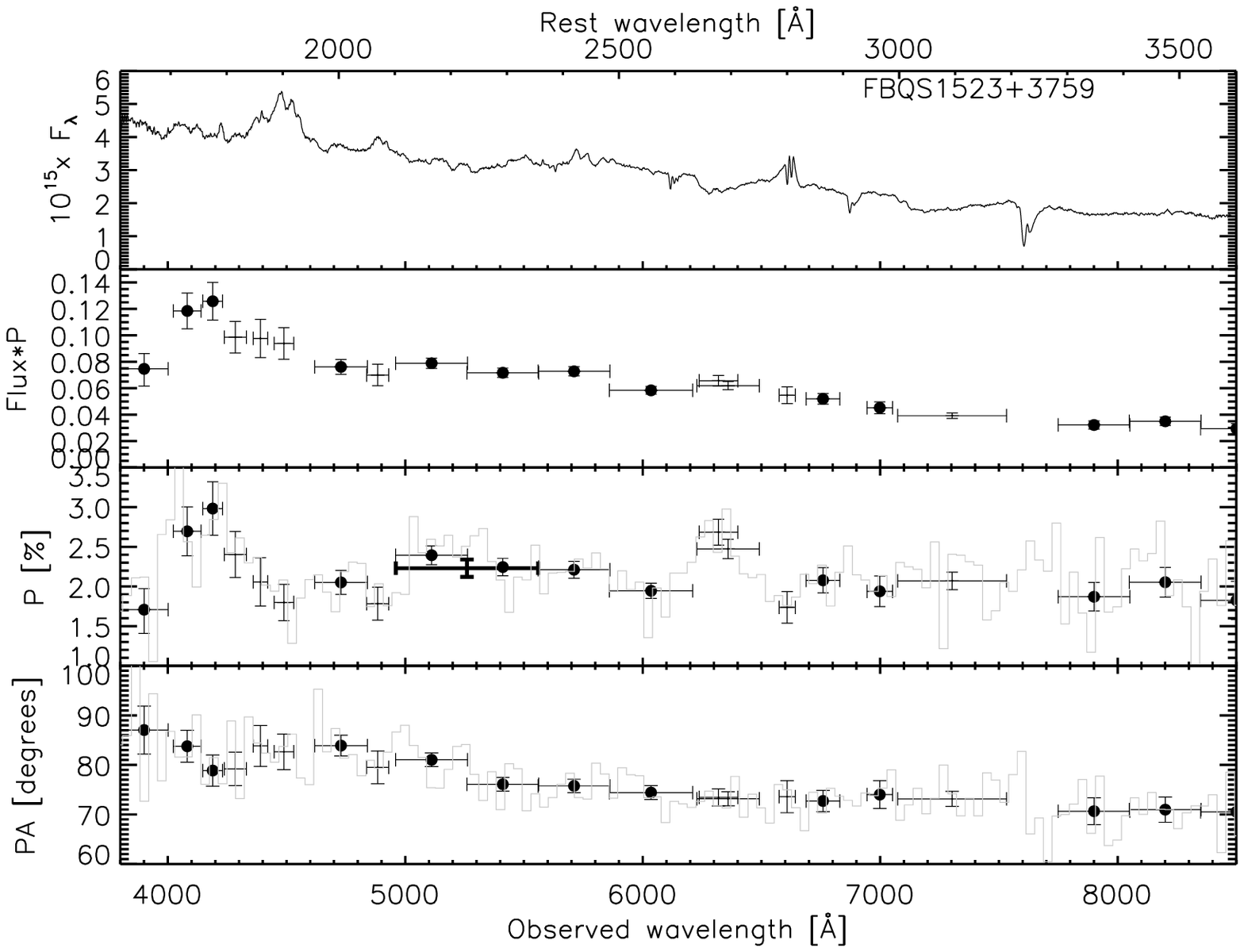}}
\end{figure}

\begin{figure}
 \ContinuedFloat
 \centering
  \subfloat[][Fig. 2$y$]{\includegraphics[width=5in]{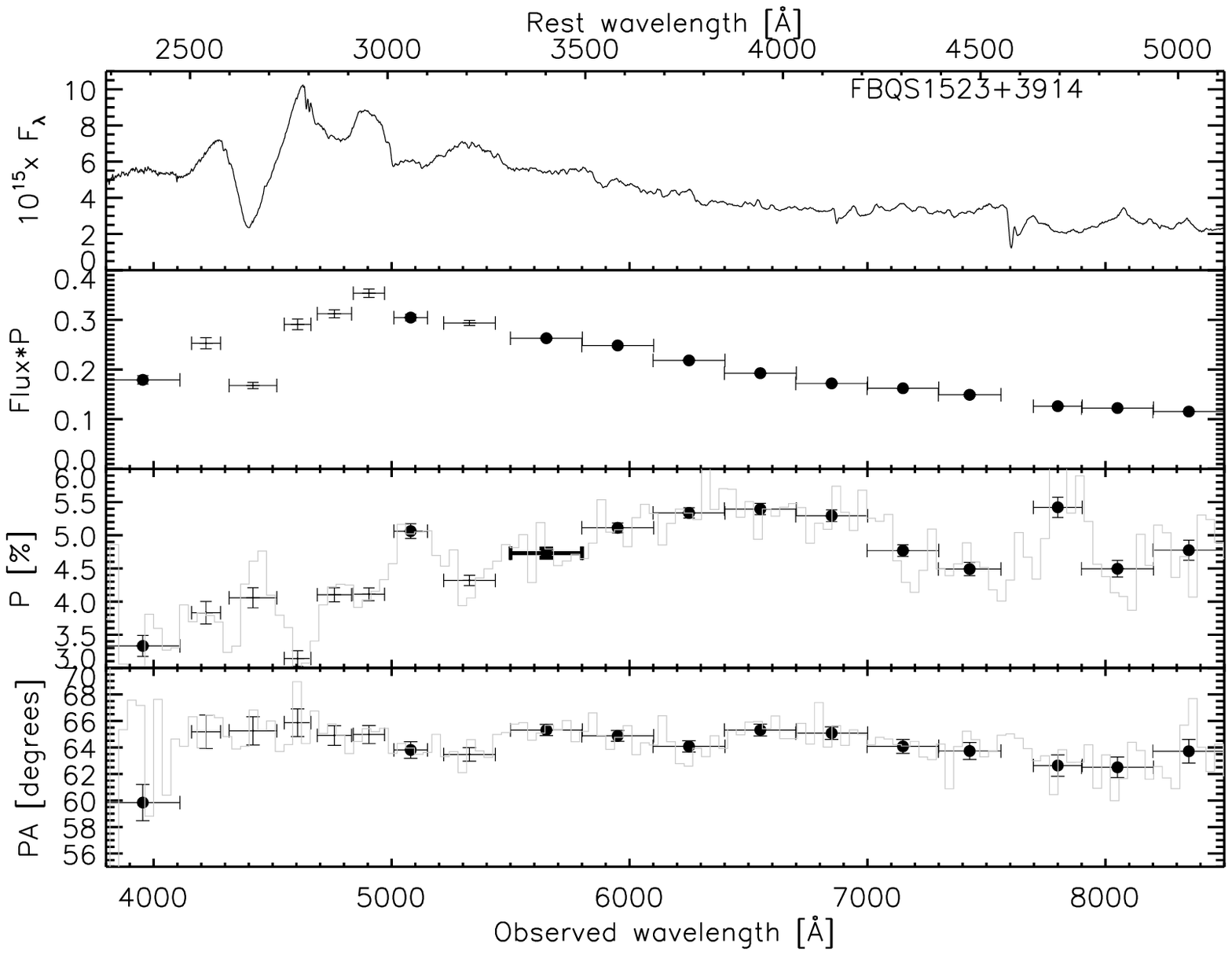}}
\end{figure}

\begin{figure}
 \ContinuedFloat
 \centering
  \subfloat[][Fig. 2$z$]{\includegraphics[width=5in]{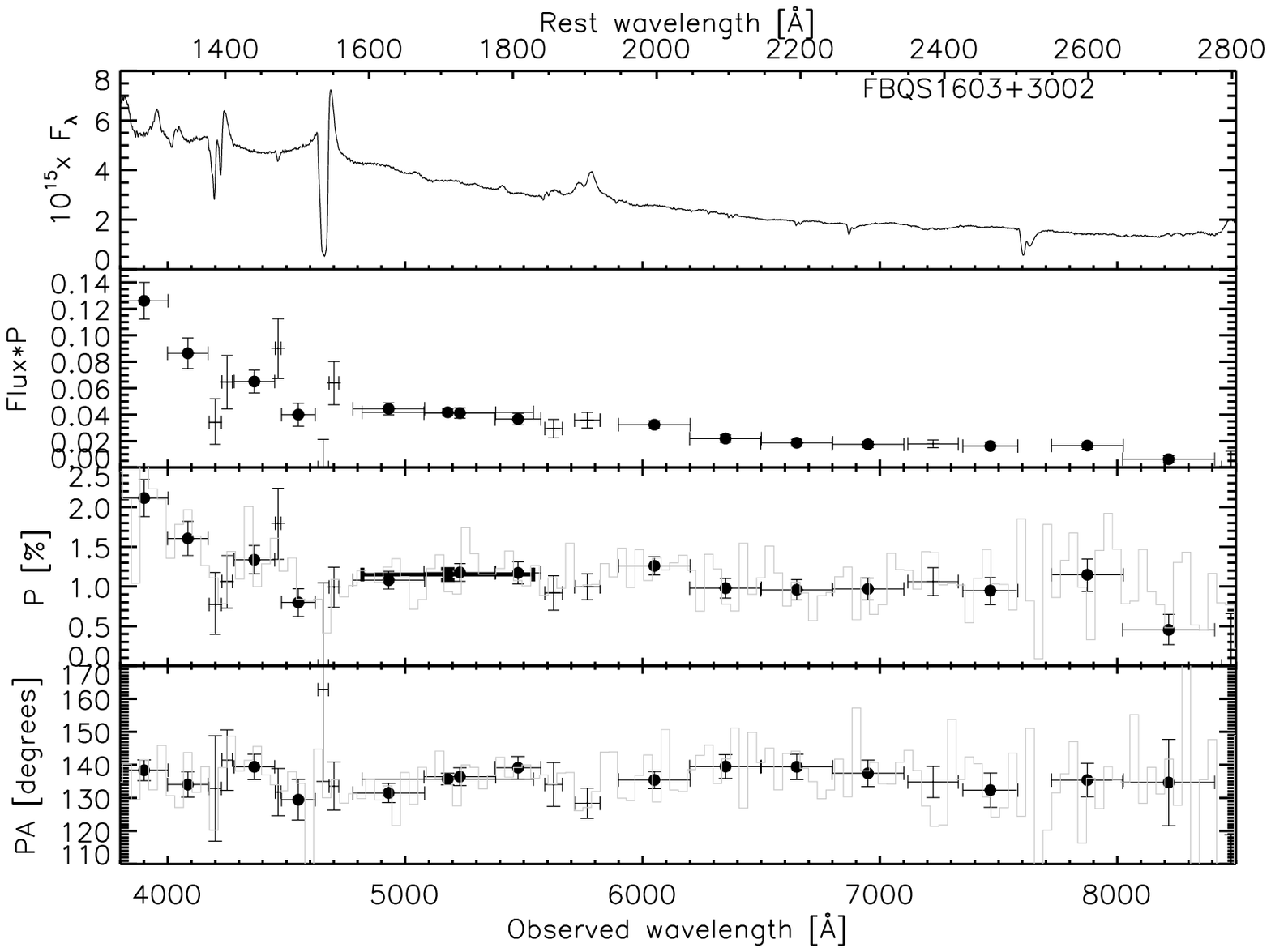}}
\end{figure}

\begin{figure}
 \ContinuedFloat
 \centering
  \subfloat[][Fig. 2$aa$]{\includegraphics[width=5in]{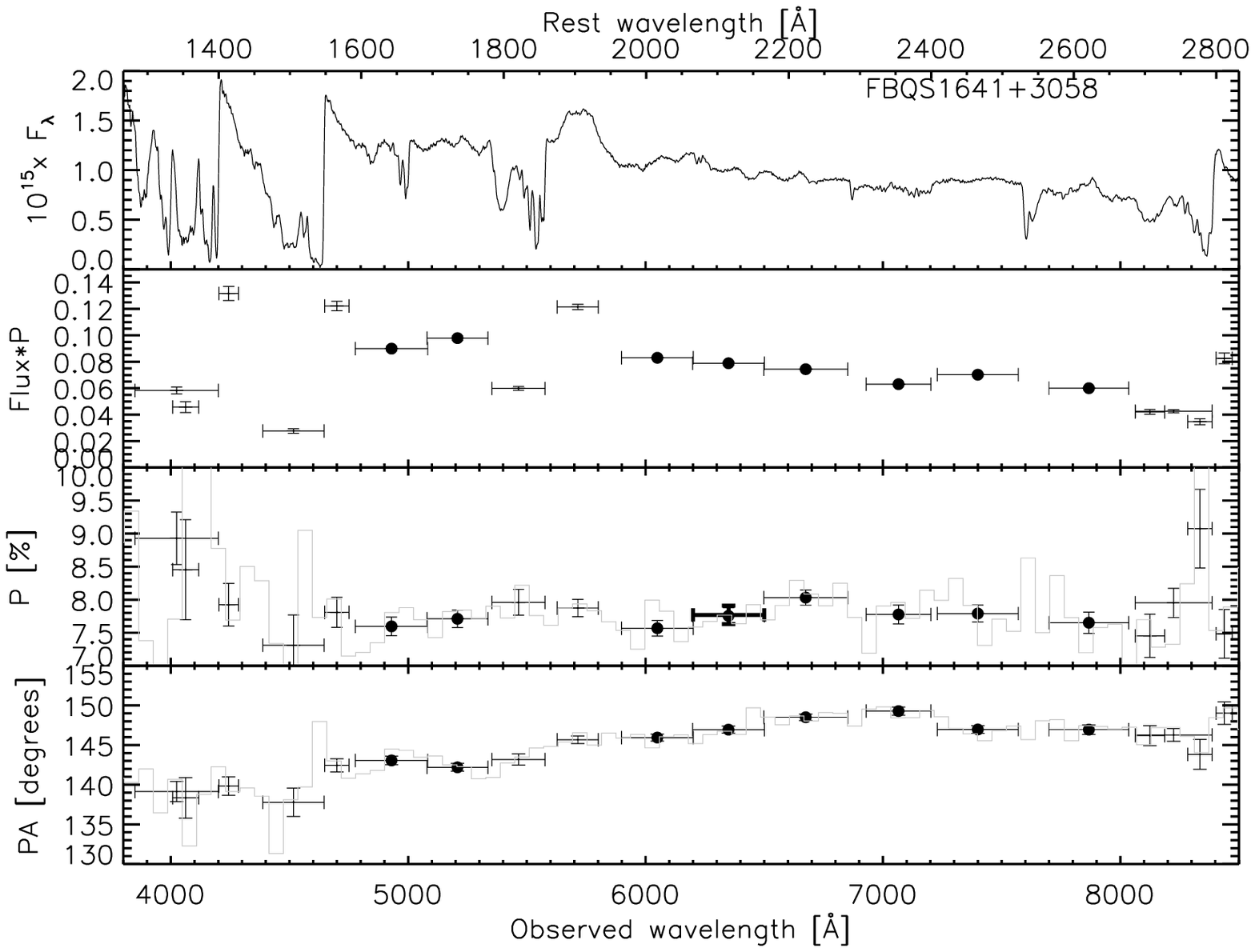}}
\end{figure}

\begin{figure}
 \ContinuedFloat
 \centering
  \subfloat[][Fig. 2$bb$]{\includegraphics[width=5in]{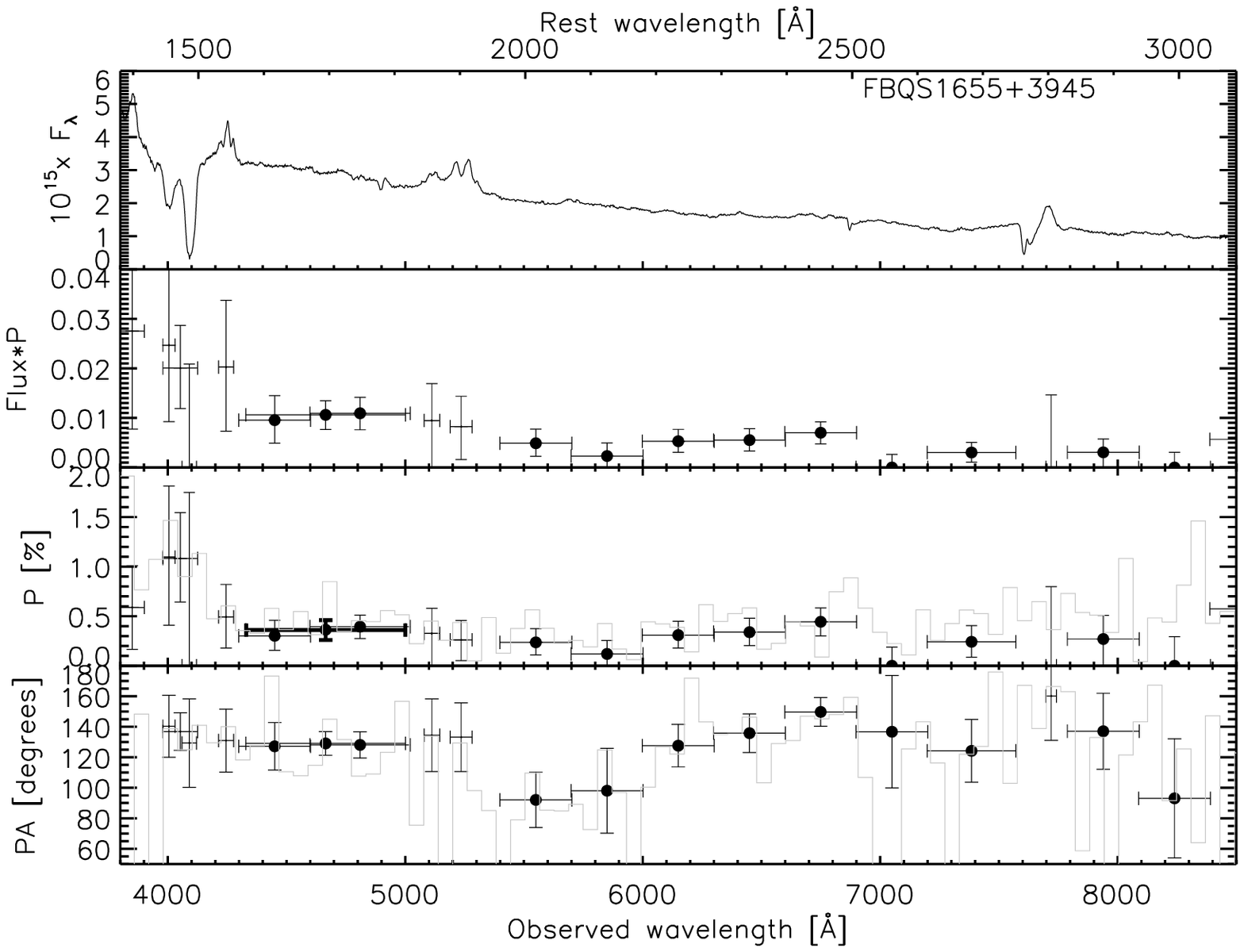}}
\end{figure}

\begin{figure}
 \ContinuedFloat
 \centering
  \subfloat[][Fig. 2$cc$]{\includegraphics[width=5in]{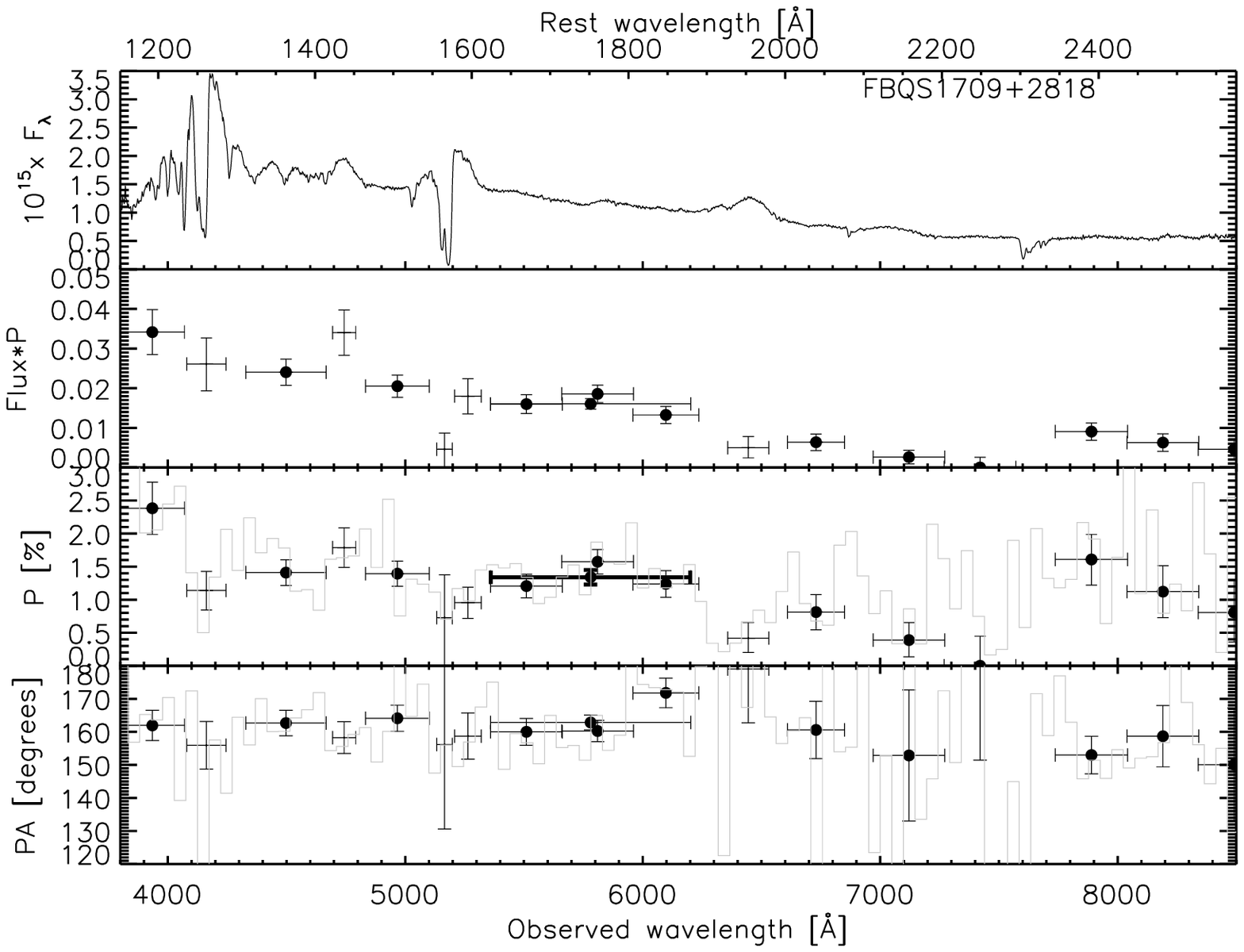}}
\end{figure}

\begin{figure}
 \ContinuedFloat
 \centering
  \subfloat[][Fig. 2$dd$]{\label{fig2ee}\includegraphics[width=5in]{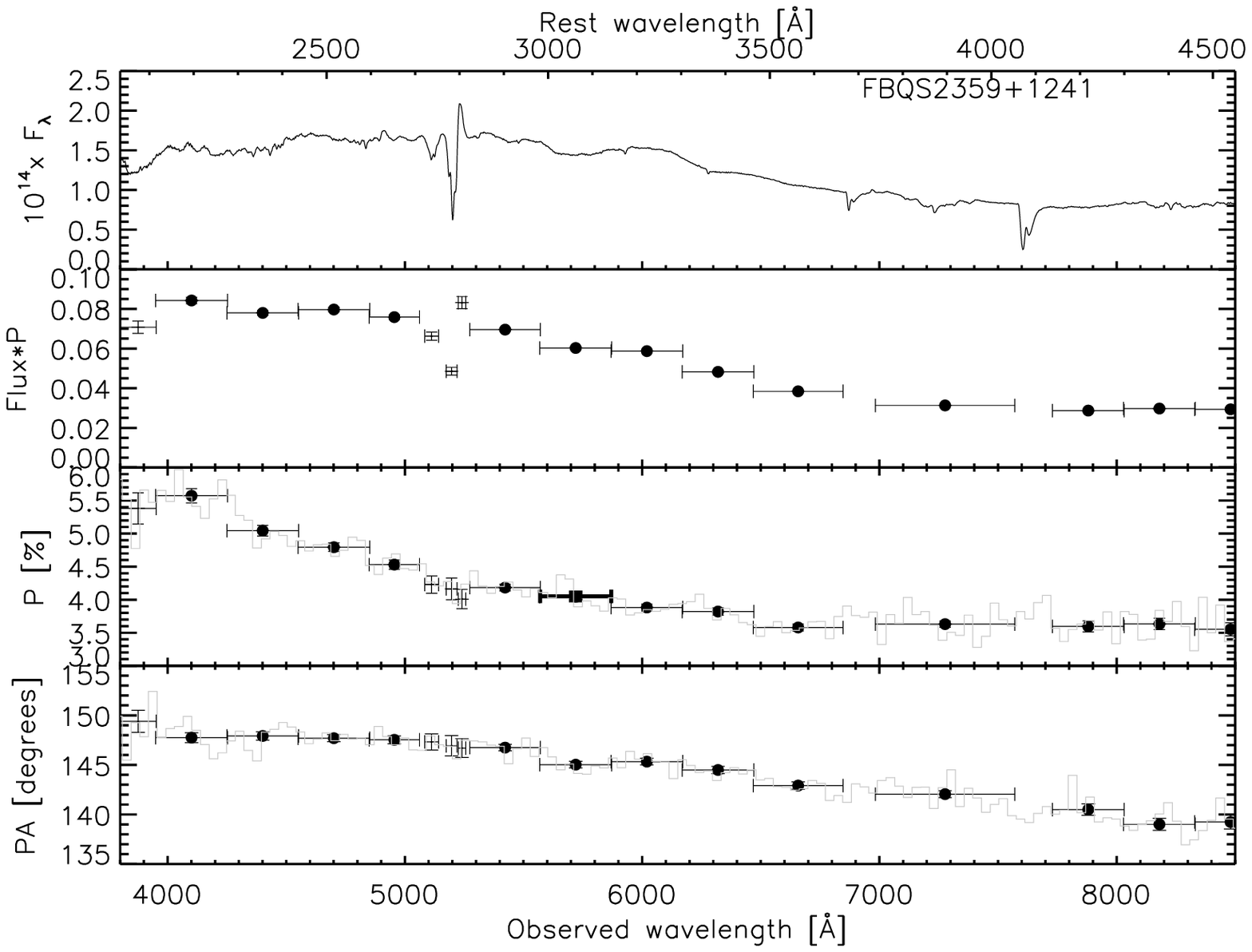}}
\end{figure}

\clearpage

\clearpage

\clearpage

\clearpage

\clearpage

\end{document}